\documentclass[aps,prb,preprint, amsmath,amssymp,showpacs]{revtex4-1}

\usepackage[T1]{fontenc}
\usepackage[latin1]{inputenc}
\usepackage[dvips]{graphicx,color}
\usepackage{rotating}
\usepackage{bm}        
\usepackage{amssymb}   
\usepackage{amsmath} 
\usepackage{bm}
\usepackage{CJK}
\def\jcp#1#2#3{J.~Chem.~Phys.~{\bf #1},\ #2\ (#3)}

\def\pra#1#2#3{Phys.~Rev.~A~{\bf #1},\ #2\ (#3)}
\def\prl#1#2#3{Phys.~Rev.~Lett.~{\bf #1},\ #2\ (#3)}

\def\k1{k_1}
\def\k2{k_2}
\def\q1{q_1}
\def\q2{q_2}

\def\({\left (}
\def\){\right )}
\def\[{\left [}
\def\]{\right ]}

\newcommand{\beq}{\begin{equation}}
\newcommand{\eeq}{\end{equation}}

\newcommand{\threejm}[6]{ \left(\begin{array}{ccc} #1 & #3 & #5\\
                                              #2 & #4 & #6
                                \end{array}
                          \right)}
\newcommand{\sixj}[6]{ \left\{\begin{array}{ccc} #1 & #3 & #5\\
                                              #2 & #4 & #6
                                \end{array}
                          \right\}}

\newcommand{\cgc}[6]{\begin{bmatrix}
{#1} & {#3} & {#5} \\
{#2} & {#4} & {#6} \\
\end{bmatrix}
}

\begin{document}
\begin{CJK}{UTF8}{gbsn}
\date{\today}
\flushbottom \draft
\title{Cold collisions of polyatomic molecular radicals with $S$-state atoms in a magnetic field: An $\bm{ ab\,\, initio}$ study of He + CH$_2(\tilde{X})$ collisions}

\author{T. V. Tscherbul and T. A. Grinev}
\affiliation{Chemical Physics Theory Group, Department of Chemistry, University of Toronto, Toronto, Ontario, M5S 3H6, Canada}\email[]{ttscherb@chem.utoronto.ca}
\author{H.-G. Yu}
\affiliation{Chemistry Department, Brookhaven National Laboratory, Upton, New York 11973} 
\author{A. Dalgarno}
\affiliation{Harvard-MIT Center for Ultracold Atoms, Cambridge, Massachusetts 02138}
\affiliation{Institute for Theoretical Atomic, Molecular and Optical Physics, 
Harvard-Smithsonian Center for Astrophysics, Cambridge, Massachusetts 02138} 
\author{Jacek K{\l}os}  \author{Lifang Ma  (马莉芳)}
\affiliation{Department of Chemistry and Biochemistry, University of Maryland, College Park, Maryland 20742}
\author{Millard H. Alexander}
\affiliation{Department of Chemistry and Biochemistry and Institute for Physical Science and Technology, University of Maryland, College Park, Maryland 20742}

\begin{abstract}
We develop a rigorous quantum mechanical theory for collisions of polyatomic molecular radicals with S-state atoms in the presence of an external magnetic field. The theory is based on a fully uncoupled space-fixed basis set representation of the multichannel scattering wavefunction. Explicit expressions are presented for the matrix elements of the scattering Hamiltonian for spin-1/2 and spin-1 polyatomic molecular radicals interacting with structureless targets. The theory is applied to calculate the cross sections and thermal  rate constants for spin relaxation in low-temperature collisions of the prototypical organic molecule methylene [CH$_2(\tilde{X}^3B_1)$] with He atoms. To this end, two highly accurate three-dimensional potential energy surfaces (PESs) of the He-CH$_2(\tilde{X}^3B_1)$ complex are developed using the state-of-the-art  CCSD(T) method and large basis sets. Both PESs exhibit shallow minima and are weakly anisotropic. Our calculations show that spin relaxation in collisions of CH$_2$, CHD, and CD$_2$ molecules with He atoms occurs at a much slower rate than elastic scattering over a large range of temperatures (1 $\mu$K -- 1 K) and magnetic fields (0.01 -- 1 T), suggesting excellent prospects for cryogenic helium buffer-gas cooling of ground-state $ortho$-CH$_2(\tilde{X}^3B_1)$ molecules in a magnetic trap.  Furthermore, we find that $ortho$-CH$_2$ undergoes collision-induced spin relaxation much more slowly than  $para$-CH$_2$, which indicates that magnetic trapping can be used to  separate  nuclear spin isomers of open-shell polyatomic molecules. 



\end{abstract}

\maketitle
\end{CJK}
\clearpage
\newpage

\section{Introduction}

Cooling and trapping polyatomic molecules is a promising research direction within the rapidly expanding field of cold molecular gases \cite{NJP}. Because of their rich internal structure and coupled vibrational degrees of freedom, polyatomic molecules offer new research opportunities at the interface between chemistry, physics, combustion \cite{Smith1}, atmospheric chemistry \cite{Smith2}, and astrochemistry \cite{HerbstARAA}. Studies of intramolecular energy redistribution in complex organic molecules yield insight into a broad range of phenomena, ranging from biochemical reactions involved in visual transduction \cite{Polli} to quantum decoherence in mesoscopic systems \cite{C60}. Unlike diatomic molecules, polyatomic molecules composed of four (or more) atoms can exist in the form of left and right-handed isomers and high-precision spectroscopy of cold molecular ensembles can provide insight into parity violation \cite{Quack} and Hund's paradox \cite{Hornberger}. Internal cooling increases the population of low-lying rovibrational states and hence the molecular response to external field perturbations, while cooling of external (translational) degrees of freedom enhances the molecule-field interaction time. High-precision spectroscopy of cold polyatomic molecules may thus enable ultrasensitive molecular detection and separation of different conformers  \cite{Meijer}  as well as the development of novel time and frequency standards \cite{Metrology}.

While indirect cooling techniques such as photo-association and magneto-association are highly efficient at producing ultracold alkali-dimers \cite{KRb}, they cannot be extended in a straightforward manner to cool polyatomic molecules due to the lack of efficient laser cooling methods for all but a handful of atomic species \cite{NJP}. Direct cooling techniques such as sympathetic cooling \cite{BufferGasCooling,BufferGasCooling2,HeNH07,HeNH}, velocity filtering \cite{Rempe}, rotating nozzle slowing \cite{nozzle}, and Stark deceleration  \cite{StarkDeceleration} are free from these difficulties, as demonstrated by recent experiments on the production of cold ensembles of fully deuterated ammonia \cite{MeijerND3}, benzonitrile \cite{MeijerBN},  formaldehyde \cite{formaldehyde}, and naphthalene \cite{Dave10}. Slow beams of polyatomic molecules thus produced have already been used, in combination with electromagnetic traps, to explore the dynamics of OH + ND$_3$ \cite{OH-ND3} and Li + SF$_6$ \cite{LiSF6} collisions at low temperatures. However, quantum collision dynamics of polyatomic molecules in the presence of external electromagnetic fields remains poorly understood,  necessitating the development of theoretical tools for {\it ab initio} modeling of these pioneering experiments as well as for the realization of efficient strategies for controlled molecular cooling \cite{NJP}. 



Theoretical studies of polyatomic molecule collisions in the absence of external fields were pioneered by Green \cite{Green}, who developed a formal quantum theory of collisions involving asymmetric tops, and carried out scattering calculations on various problems of astrophysical interest, including He + HCN, He + H$_2$O, and He + NH$_3$ \cite{Green,GreenH2O,GreenHCN}. Many authors employed Green's formalism to carry out accurate scattering calculations of rotational cooling rates for interstellar molecules colliding with atomic and molecular species based on accurate {\it ab initio} interaction potentials \cite{Faure}. Several theoretical studies have used a combination of high-level {\it ab initio} and quantum scattering calculations to explore the dynamics of collision-induced rovibrational relaxation of methylene (CH$_2(a,\tilde{X})$) and methyl (CH$_3$) radicals in room-temperature He buffer gas \cite{Millard,MillardCH2new}, for which accurate rate measurements have recently become available \cite{Greg}.

 A number of related theoretical studies have focused on field-free collisions of H$_2$O, CO$_2$, and benzene molecules with He atoms at temperatures below 1 K \cite{Stancil, Barker}. These calculations were motivated by the ongoing experiments on sympathetic cooling of optically decelerated \cite{OD} polyatomic molecules with rare gas atoms \cite{Barker}. In particular, evidence was reported for slow rotational energy transfer in cold collisions of highly rotationally excited CO$_2$ molecules with He atoms \cite{Rotors}. A recent classical trajectory study of He-naphthalene scattering \cite{Zhiying} has shown that the unexpected lack of cluster formation observed in buffer-gas cooling experiments on naphthalene \cite{Dave10} can be attributed to the short lifetime of the He-naphthalene complex.

We have recently presented a rigorous quantum scattering methodology for numerical calculations of collisional properties of polyatomic molecular radicals with $S$-state atoms in the presence of an external magnetic field \cite{prl11}. The method is based on the fully uncoupled space-fixed representation of the scattering wave function originally introduced by Volpi and Bohn  \cite{VolpiBohn} and Krems and Dalgarno \cite{Roman04} to investigate collisions of diatomic molecules in a magnetic field. In Ref. \citenum{prl11} we studied the quantum dynamics of spin relaxation in collisions of spin-1/2 and spin-1 polyatomic molecular radicals induced by collisions with He atoms. We found that many of these polyatomic molecules  have favorable collisional properties (that is, large ratios of elastic to inelastic collisions rates) with He atoms, making them amenable to cryogenic buffer-gas cooling and possibly sympathetic cooling with $S$-state atoms in a magnetic trap \cite{prl11}. 

Here, we present a full account of the theory and computational methodology outlined in our previous work \cite{prl11}. We begin with the derivation of coupled-channel quantum scattering equations for a polyatomic molecule colliding with an $S$-state atom in the presence of an external magnetic field (Sec. IIB). In Sec. III, we use the newly developed scattering methodology to address the question of collisional stability of methylene radicals in cold $^3$He gas in the presence of an external magnetic field. After a brief outline of the {\it ab initio} techniques employed to obtain the adiabatic PES for the He-CH$_2$ complex (Sec. IIIA), we present and discuss the main results for elastic and inelastic cross sections and thermal rate constants for spin relaxation in  He + CH$_2$ collisions (Sec. IIIB). We then analyze the scaling properties of the inelastic cross sections and discuss the mechanism of spin relaxation of different nuclear spin isomers (Sec. IIIC) and isotopomers (Sec. IIID) of CH$_2$. Section IV summarizes the main results of this work.

Atomic units are used throughout unless stated otherwise.

\section{Theory}

In this section, we will outline our approach to the solution of the atom -- polyatomic molecule collision problem in the presence of an external magnetic field. We first describe our {\it ab initio} calculations of the potential energy surfaces (PESs) of the He-CH$_2$ complex. In Sections IIB and IIC, we generalize this work to collisions of polyatomic molecules (asymmetric tops) with $S$-state atoms. We present the derivation of the matrix elements required to parametrize the close-coupling (CC) scattering equations for collisions of polyatomic molecular radicals with $S$-state atoms. Sec. II C presents a brief description of our numerical calculations.


\subsection{Hamiltonian}\label{Hamiltonian}

The Hamiltonian for an open-shell polyatomic molecule colliding with a $^1S_0$ atom in the presence of an external magnetic field may be written as
\begin{equation}\label{H}
\hat{H} = -\frac{1}{2\mu R}\frac{\partial^2}{\partial R^2}R + \frac{\hat{l}^2}{2\mu R^2} + V(\bm{R},\hat{\Omega}) + \hat{H}_\text{mol},
\end{equation}
where $\mu$ is the reduced mass of the collision complex, $\bm{R}=R\hat{R}$ is the atom-molecule separation vector of length $R$, $\hat{l}^2$ is the orbital angular momentum for the collision, $V(\bm{R},\hat{\Omega})$ is the anisotropic potential describing the atom-molecule interaction, and $\hat{H}_\text{mol}$ is the asymptotic Hamiltonian describing the internal structure of the molecule and its interaction with external fields. In this work, we focus on collisions at low incident kinetic energies (10$^{-4}$ -- 100 K), much lower than the typical energy scale for intramolecular electronic and vibrational excitations (typically, hundreds and thousands of Kelvin) so we can assume that the molecule remains in its ground electronic and vibrational states during the collision.  Thus we can use the following  effective Hamiltonian \cite{Bunker,CHD,CD2} to describe the rotational and fine structure of a polyatomic molecular radical like CH$_2$ \cite{note1}:
\begin{equation}\label{Hmol}
\hat{H}_\text{mol} = \hat{H}_\text{rot} + \hat{H}_\text{cd}  + \hat{H}_\text{sr} + \hat{H}_\text{ss} + \hat{H}_\text{ext}.
\end{equation}
The rotational part of this Hamiltonian is that of a rigid asymmetric top with rotational constants $A$, $B$, and $C$: $ \hat{H}_\text{rot} = A\hat{N}_a^2 +  C\hat{N}_b^2 + B\hat{N}_c^2$, where the operators $\hat{N}_\alpha$ ($\alpha = a,b,c$) yield the molecule-fixed (MF) components of rotational angular momentum $\hat{\bm{N}}$. The fine-structure Hamiltonians $\hat{H}_\text{sr}$ and $\hat{H}_\text{ss}$ describe the spin-rotation and spin-spin interactions in polyatomic molecules with non-zero total spin $S$ (radicals). The term $\hat{H}_\text{cd}$ describes the centrifugal distortion of rotational levels and the term $\hat{H}_\text{ext}$ describes the interaction of the molecular spin with an external magnetic field.
 
 Here, we choose the MF axes to correspond to the symmetry axis  of the molecule, as illustrated in Fig. \ref{fig:MF}. This choice is the same as adopted by Hutson in his theoretical study of Ar + H$_2$O collisions \cite{Hutson} but differs from the recent work of Dagdigian and co-workers, where the principal axis of the triatomic is used to define the $z$ axis~\cite{PaulD_H2O,Millard}. The CH$_2$ molecule lies in the $xz$ plane, with the $z$-axis bisecting the HCH angle, and the $y$-axis perpendicular to the molecular plane. With this choice, the He-CH$_2$ interaction potential is symmetric under the reflection in the $xz$ plane. The rotational part of the effective Hamiltonian (\ref{Hmol}) becomes
\begin{equation}\label{Hrot}
\hat{H}_\text{rot} =  A\hat{N}_x^2 +  C\hat{N}_y^2 + B\hat{N}_z^2,
\end{equation}
and the centrifugal distortion Hamiltonian is (up to fourth order) \cite{Bunker}
\begin{equation}\label{Hcd}
\hat{H}_\text{cd} = -\Delta_N \hat{N}^4 - \Delta_{NK} \hat{N}^2 \hat{N}_x^2 - \Delta_K \hat{N}_x^4 - 2\delta_N \hat{N}^2 (\hat{N}_z^2 - \hat{N}_y^2).
\end{equation}

The spin-rotation interaction is given by \cite{Bunker,BunkerBook}
\begin{equation}\label{Hsr_BF}
\hat{H}_\text{sr} = \gamma_x \hat{N}_x \hat{S}_x  + \gamma_y \hat{N}_y \hat{S}_y + \gamma_z \hat{N}_z \hat{S}_z,
\end{equation}
where $\hat{\bm{S}}$ is the electron spin, $\gamma_\alpha$ are the spin-rotation constants. Eq. (\ref{Hsr_BF}) gives the spin-rotation interaction in terms of the MF angular momentum operators $\hat{\bm{N}}$ and $\hat{\bm{S}}$. The corresponding expression in terms of SF angular momentum operators is derived in Appendix A.


 The spin-spin interaction may be written as 
\begin{equation}\label{Hss_BF}
\hat{H}_\text{ss} 
=  \sum_{p=-2}^2 \biggl{[} \frac{1}{2} (E+D) [\mathcal{D}^2_{p,2}(\hat{\Omega}) + \mathcal{D}^2_{p,-2}(\hat{\Omega})] \\ +   \frac{3E-D}{\sqrt{6}} \mathcal{D}^2_{p,0}(\hat{\Omega}) \biggr{]} [\hat{\bm{S}}\otimes\hat{\bm{S}}]^{(2)}_p,
\end{equation}
where $\mathcal{D}^2_{p,p'}(\hat{\Omega})$ are the Wigner $D$-functions, $\hat{\Omega}$ are the Euler angles which specify the orientation of MF axes in the space-fixed (SF) coordinate frame, and $D$ and $E$ are the zero-field-splitting parameters \cite{Bunker,BunkerBook}. Equation (\ref{Hss_BF}) may be derived from the standard textbook expression for the spin-spin interaction in the MF frame  \cite{Raynes,Bunker,BunkerBook} as described in Appendix A.  
Note that the spin-spin interaction is only different from zero in molecular radicals bearing more than one unpaired electron ($S=1$).

The interaction of the polyatomic molecule with an external magnetic field is described by the Hamiltonian \cite{Bunker}
\begin{equation}\label{Hext}
\hat{H}_\text{ext} = \mu_0 \bm{B} \cdot \mathbf{g}_s \cdot \hat{\bm{S}},
\end{equation}
where $\bm{B}$ is the magnetic field vector, $\mu_0$ is the Bohr magneton, and $\mathbf{g}_s$ is the $g$-tensor. Following Ref. \cite{Bunker}, we take $\mathbf{g}_s = 2 \mathbf{I}$ where $\mathbf{I}$ is the unit matrix, so Eq. (\ref{Hext}) reduces to a form commonly used for diatomic $^3\Sigma$ molecules
\begin{equation}\label{Hext2}
\hat{H}_\text{ext} = 2 \mu_0 B \hat{S}_Z,
\end{equation}
where $\hat{S}_Z$ yields the projection of $\hat{\bm{S}}$ on the magnetic field axis. The rotational and nuclear spin Zeeman interactions are much smaller than the $\hat{H}_\text{ext}$ term, and we neglect them in this work. Table II lists the molecular constants used in this work to parametrize the Hamiltonian (\ref{Hmol}) for CH$_2(\tilde{X})$.


\subsection{Coupled-channel equations, S-matrix elements, and cross sections}

To solve the quantum scattering problem for the Hamiltonian (\ref{H}), we expand the wave function of the collision complex in direct products of internal basis functions $|\alpha\rangle$ and partial waves of relative motion $| lm_l\rangle$ in the space-fixed (SF) coordinate frame
\begin{equation}\label{expansion}
| \Psi \rangle = \frac{1}{R} \sum_{\alpha,l,m_l} F_{\alpha lm_l}(R) |\alpha\rangle |lm_l\rangle,
\end{equation}
where the basis functions $|\alpha\rangle = |NM_NK_N\rangle |SM_S\rangle$ describe the rotational and spin degrees of freedom of the polyatomic molecule, $|NM_NK_N\rangle = \sqrt{(2N+1)/8\pi^2} \mathcal{D}^{N*}_{M_N,K_N}(\hat{\Omega})$ is a symmetric-top eigenfunction, 
$M_N$ ($K_N$) is the projection of $\hat{\bm{N}}$ on the SF (MF) quantization axes, and $M_S$ is the SF projection of $\hat{\bm{S}}$.  We note that the projection $M=M_N+M_S+m_l$ of the total angular momentum on the SF quantization axis is rigorously conserved for collisions in a magnetic field. As in the atom-diatom case considered previously \cite{Roman04} this conservation can be exploited to factorize the collision problem into smaller problems which can be solved independently for each $M$.

Using the expansion (\ref{expansion}) in combination with the time-independent Schr{\"o}dinger equation for Hamiltonian (\ref{H}) leads to a system of close-coupling (CC) equations 
\begin{equation}\label{CC}
\left[ \frac{d^2}{dR^2} - \frac{l(l+1)}{R^2} + 2\mu E \right] F_{\alpha lm_l}(R) =  2 \mu \sum_{\alpha',l',m_l'}  \langle  \alpha | \langle l m_l | V(R,\theta,\phi) + \hat{H}_\text{mol}| \alpha' \rangle | l' m_l'\rangle F_{\alpha' l'm_l'}(R).
\end{equation}
The matrix elements of the asymptotic Hamiltonian $\hat{H}_\text{mol}$ and of the atom-molecule interaction potential $V(R,\theta,\phi)$ are evaluated as described below. As in the atom-diatom case \cite{Roman04},  the matrix of the asymptotic Hamiltonian is not diagonal in the fully uncoupled SF representation. In order to properly apply the scattering boundary conditions, we introduce an asymptotic basis \cite{Roman04}
\begin{equation}\label{basis_gamma}
|\gamma\rangle |lm_l\rangle =  | lm_l\rangle \sum_{\alpha} C_{\alpha \gamma} |\alpha\rangle,
\end{equation}
which has the property
\begin{equation}\label{Hmol_gamma_gammaprime}
\langle \gamma | \hat{H}_\text{mol} | \gamma'\rangle = \delta_{\gamma\gamma'}\epsilon_\gamma,
\end{equation}
where $\epsilon_\gamma$ are the asymptotic eigenvalues, which define scattering channels in the presence of an external magnetic field. The coefficients $C_{\alpha\gamma}$ in Eq. (\ref{basis_gamma}) are independent of $l$ and $m_l$ and form the matrix $\mathbf{C}$, which satisfies $\mathbf{C}^T \mathbf{H}_\text{mol} \mathbf{C} = \mathbf{E}$, where $\mathbf{E}$ is the diagonal matrix of eigenvalues $\epsilon_\gamma$. The asymptotic behavior of the solutions to CC equations (\ref{CC}) in the asymptotic basis \cite{Roman04} defines the elements of the scattering $S$-matrix
\begin{equation}\label{boundary_conditions}
F^M_{\gamma' l'm_l'}(R)  \simeq \delta_{\gamma\gamma'} \delta_{ll'}\delta_{m_l m_l'} e^{-i(k_\gamma R-\pi l/2)} - \left( \frac{k_\gamma}{k_{\gamma'}} \right)^{1/2} S^M_{\gamma lm_l; \gamma' l' m_l'} e^{i(k_\gamma R-\pi l/2)} \quad (R\to \infty),
\end{equation}
as in the atom-diatom case \cite{Roman04}. The cross sections for the collision-induced transition $\gamma\to \gamma'$ in the polyatomic molecule may be expressed through the $S$-matrix elements as
\begin{equation}\label{x_sections}
\sigma_{\gamma \to \gamma'} (E_C) = \frac{\pi}{k_\gamma^2} \sum_M \sum_{l,\,m_l} \sum_{l',\, m_l'} | \delta_{\gamma\gamma'} \delta_{ll'}\delta_{m_l m_l'} - S^M_{\gamma lm_l; \gamma' l' m_l'} (E_C) |^2,
\end{equation}
where $k_\gamma^2 = 2\mu (E-\epsilon_\gamma)$ is the wavevector for channel $\gamma$, $E=E_C+\epsilon_\gamma$ is the total energy, and $E_C$ is the  collision energy.




\subsection{Matrix elements in the fully uncoupled space-fixed basis}

The matrix elements entering the CC equations (\ref{CC}) can be expressed analytically via molecular constants and expansion coefficients of the atom-molecule interaction potential. 
In this section, we present the expressions for these matrix elements in the fully uncoupled SF basis of Eq. (\ref{expansion}). 

We start with the asymptotic Hamiltonian of the polyatomic molecule (\ref{Hmol}). First, we note that since the asymptotic Hamiltonian (\ref{Hmol}) is independent of the atom-molecule orientation, its matrix elements are  independent of $l$ and $m_l$:
\begin{equation}\label{diagonal}
\langle  \alpha | \langle l m_l | \hat{H}_\text{mol}| \alpha'\rangle | l' m_l'\rangle = \delta_{ll'}\delta_{m_l m_l'} \langle \alpha | \hat{H}_\text{mol} | \alpha' \rangle.
\end{equation}
The matrix element on the right-hand side can be written as the sum of rotational (\ref{Hrot}), centrifugal distortion (\ref{Hcd}), fine-structure (\ref{Hsr_BF}), (\ref{Hss_BF}) and magnetic field-dependent terms (\ref{Hext2}) given by Eq. (\ref{Hmol}).


 In order to evaluate the matrix elements of $\hat{H}_\text{rot}$, we express the MF angular momentum operators $\hat{N}_x$ and $\hat{N}_y$ in terms of the ladder operators $\hat{N}_\pm$ and use the well-known expressions for their matrix elements \cite{Zare} to obtain 
\begin{multline}\label{me_Hrot}
\langle N M_N K_N | \langle SM_S  | \hat{H}_\text{rot}  | N' M_N' K_N' \rangle |S M_S'\rangle \\ = \delta_{M_S M_S'} \delta_{NN'}\delta_{M_NM_N'}  \biggl{[} \frac{1}{4}(A-C)\lambda_\pm (N,K_N')\lambda_\pm(N,K_N'\pm1) \delta_{K_N,K_N'\pm 2} \\ +  \frac{1}{2}(A+C) [N(N+1) - K_N^2] \delta_{K_N K_N'} +  BK_N^2\delta_{K_NK_N'} \biggr{]},
\end{multline}
where
\begin{equation}
\lambda_\pm (N,K_N) = [N(N+1)-K_N(K_N\pm1)]^{1/2}.
\end{equation}
The matrix elements of the centrifugal distortion term (\ref{Hcd}) can be evaluated following the same procedure. Explicit expressions for the matrix elements of $\hat{H}_\text{cd}$ are presented in Appendix~B.

A triatomic molecule with two identical H nuclei (like CH$_2$) can exist in two different nuclear spin modifications, ortho ($o$) and para ($p$). Nuclear spin selection rules strongly suppress collision-induced transitions between the energy levels of different nuclear spin isomers  \cite{BunkerBook,Chapovsky}. 
Because the Hamiltonian is invariant to interchange of the two H atoms, $o$-CH$_2$ and $p$-CH$_2$ can be considered separate molecular species \cite{GreenH2O}.
 We obtain the energy level spectrum of each nuclear spin isomer by diagonalizing the Hamiltonian (\ref{Hmol}) using a restricted basis of symmetric-top eigenfunctions $|NM_NK_N\rangle$ with $(-1)^{K_N}=1$ (for $o$-CH$_2$) and $(-1)^{K_N}=-1$ (for $p$-CH$_2$).


In order to evaluate the matrix elements of the spin-rotation interaction (\ref{Hsr_BF}), we rewrite it in the form
\begin{align}\label{Hsr_1}\notag
\hat{H}_\text{sr} & = \frac{1}{2} \biggl{[} (\gamma_x-\gamma_y) \hat{N}_{+1} - (\gamma_x+\gamma_y) \hat{N}_{-1} \biggr{]}\hat{S}_{+1} 
\\ &+ \frac{1}{2} \biggl{[} (\gamma_x-\gamma_y) \hat{N}_{-1} - (\gamma_x+\gamma_y) \hat{N}_{+1} \biggr{]}\hat{S}_{-1} + \gamma_z \hat{N}_0 \hat{S}_0,
\end{align}
where $\hat{N}_q$ and $\hat{S}_q$ are the spherical tensor components of $\hat{N}$ and $\hat{S}$ in the MF frame. Rewriting the MF spin operators in terms of their SF counterparts \cite{Zare} (see Eq. (\ref{A2}) in Appendix A), 
evaluating the integral over three Wigner $D$-functions, and rearranging the result, we obtain 
\begin{multline}\label{me_Hsr_1}
\langle N M_N K_N | \langle SM_S  | \hat{H}_\text{sr}  | N' M_N' K_N' \rangle |S M_S'\rangle  =  \ [(2N+1)(2N'+1)]^{1/2}[(2S+1)S(S+1)]^{1/2} \\ \times (-1)^{M_N'-K_N'+S-M_S}  \mathcal{M}_{NK_N;N'K_N'}
\sum_q \threejm{S}{-M_S}{1}{q}{S}{M_S'} \threejm{N}{M_N}{1}{q}{N'}{-M_N'},
\end{multline}
where 
\begin{align}\label{me_Hsr_2}
\mathcal{M}_{NK_N;N'K_N'} =  \frac{1}{2\sqrt{2}}  \biggl{[} &\lambda_- (N',K_N') (\gamma_x-\gamma_y) \threejm{N}{K_N}{1}{1}{N'}{-K_N'+1} \\ \notag
+ &\lambda_+(N',K_N') (\gamma_x+\gamma_y)  \threejm{N}{K_N}{1}{1}{N'}{-K_N'-1} \biggr{]} \\ \notag
- \frac{1}{2\sqrt{2}}  \biggl{[} &\lambda_+ (N',K_N') (\gamma_x-\gamma_y) \threejm{N}{K_N}{1}{-1}{N'}{-K_N'-1} \\ \notag
+ &\lambda_-(N',K_N') (\gamma_x+\gamma_y)  \threejm{N}{K_N}{1}{-1}{N'}{-K_N'+1} \biggr{]} +\gamma_z K_{N}'\threejm{N}{K_N}{1}{0}{N'}{-K_N'},
\end{align}
and the quantities in parentheses denote 3-$j$ symbols.
An alternative expression for the matrix elements of the spin-rotation interaction may be obtained by directly evaluating the matrix elements of the tensor product in Eq. (\ref{Hsr_BF}) in the SF frame (see Appendix A). The resulting expression is rather cumbersome, and we do not present it here. We used this expression to cross-check Eq. (\ref{me_Hsr_2}) and verified that both expressions yield the same results. 

The matrix elements of the spin-spin interaction (\ref{Hss_BF}) in the fully uncoupled basis factorize into products of matrix elements of the form
\begin{equation}\label{me_product}
\langle N M_N K_N | \mathcal{D}^2_{p,2} (\hat{\Omega}) | N' M_N' K_N' \rangle \langle SM_S | [\bm{S} \otimes \bm{S}]^{(2)}_p |S M_S'\rangle.
\end{equation}
The first matrix element reduces to a product of two 3-$j$ symbols \cite{Zare} and the second matrix element is given by Eq. (19) of Ref. \cite{Roman04}. Combining these results, we obtain (neglecting constant $S$-dependent terms)
\begin{multline}\label{me_Hss}
\langle N M_N K_N | \langle SM_S  | \hat{H}_\text{ss}  | N' M_N' K_N' \rangle |S M_S'\rangle =  [(2N+1)(2N'+1)]^{1/2} (-1)^{M_N'-K_N'+S-M_S} \\ \times \biggl{[} \frac{1}{2}(E+D) \biggl{[} \threejm{N}{K_N}{2}{2}{N'}{K_N'} + \threejm{N}{K_N}{2}{-2}{N'}{-K_N'} \biggr{]} + \frac{3E-D}{\sqrt{6}} \threejm{N}{K_N}{2}{0}{N'}{-K_N'} \biggr{]} \\
 \times \sqrt{5} [(2S+1)S(S+1)] \sixj{1}{S}{1}{S}{2}{S}
 \sum_q \threejm{N}{M_N}{2}{q}{N'}{-M_N'}  \threejm{S}{-M_S}{2}{q}{S}{M_S'}.
\end{multline}
Substituting $K_N=0$ in Eq. (\ref{me_Hss}), we recover the expression for the matrix element of the spin-spin interaction derived previously for linear $^3\Sigma$ molecules \cite{Roman04}. 

\begin{figure}[t]
	\centering
	\includegraphics[width=0.7\textwidth, trim = 0 0 0 0]{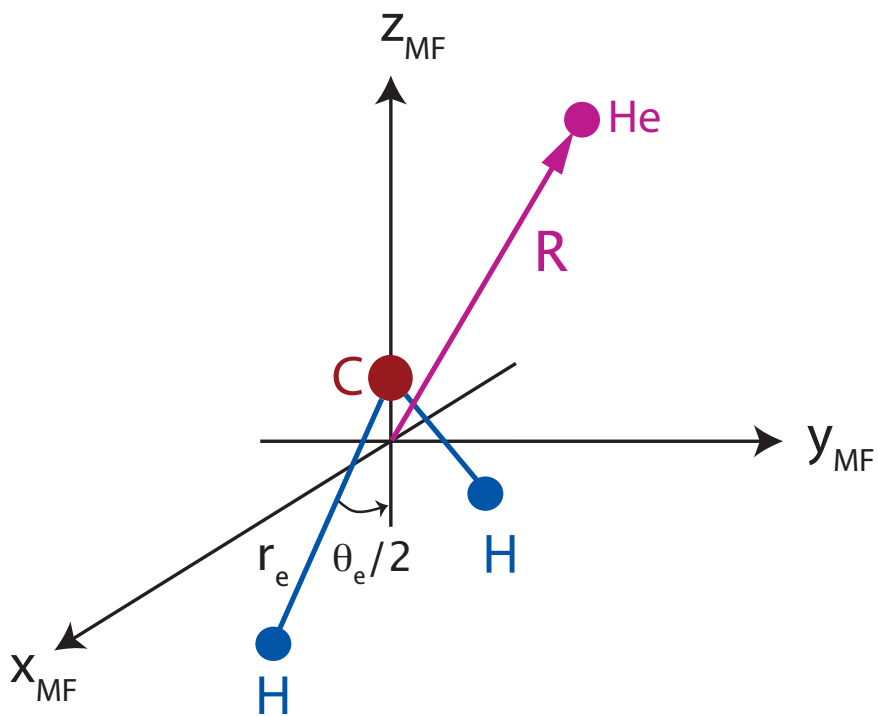}
	\renewcommand{\figurename}{Fig.}
	\caption{MF coordinate system used to describe the He-CH$_2$ interaction.}\label{fig:MF}
\end{figure}

In order to evaluate the matrix elements of the atom-molecule interaction potential in the fully uncoupled SF basis (\ref{expansion}), we expand the potential in renormalized spherical harmonics \cite{Green,Hutson}
\begin{equation}\label{Vbf}
V(R,\theta,\phi) = \sum_{\lambda = 0}^{\lambda_\text{max}} \sum_{\mu \ge 0}^\lambda V_{\lambda\mu} (R) \frac{1}{1+\delta_{\mu 0}}   [C_{\lambda\mu}(\theta,\phi) + (-1)^\mu C_{\lambda,-\mu}(\theta,\phi)],
\end{equation}
where $\theta$ and $\phi$ are the spherical polar angles of $\bm{R}$ in the MF frame (see Fig. \ref{fig:MF}) and $C_{\lambda\mu}(\theta,\phi)=[4\pi/(2\lambda+1)]^{1/2}Y_{\lambda\mu}(\theta,\phi)$. The expansion (\ref{Vbf}) is valid for any polyatomic molecule interacting with an $S$-state atom provided that $V(R,\theta,\phi)$ is symmetric under reflection in the $xz$ plane, i.e., $V(R,\theta,\phi)=V(R,\theta,-\phi)$. In our case, because of the $C_{2v}$ symmetry of CH$_2$, the potential is also symmetric under reflection in the $yz$ plane, so we have $V(R,\theta,\phi)=V(R,\theta,\pi-\phi)$. This condition implies that the expansion coefficients $V_{\lambda\mu}$ vanish for odd $\mu$. We note that this property is a consequence of our using a specific MF system shown in Fig. \ref{fig:MF}. As mentioned above,  Dagdigian and co-workers chose their MF $z$-axis to be perpendicular to the C$_2$ axis of CH$_2$ \cite{PaulD_H2O,Millard,MillardCH2new}. As a result, their $V_{\lambda\mu}$ coefficients  are not the same as defined in this work, and they do not vanish for odd $\mu$ \cite{PaulD_H2O,Millard,MillardCH2new}.

Because our SF basis functions do not depend explicitly on $\theta$ and $\phi$, we transform the expansion (\ref{Vbf}) to the SF frame. Transforming the spherical harmonics to the SF system, we obtain \cite{Green} 
\begin{equation}\label{Vsf}
V(\bm{R},\hat{\Omega}) = \sum_{\lambda = 0}^{\lambda_\text{max}} \sum_{\mu \ge 0}^\lambda V_{\lambda\mu} (R) \frac{1}{1+\delta_{\mu 0}}  \sum_{m_\lambda} [\mathcal{D}^\lambda_{m_\lambda, \mu}(\hat{\Omega}) + (-1)^\mu \mathcal{D}^\lambda_{m_\lambda,-\mu}(\hat{\Omega})] C_{\lambda m_\lambda}(\hat{R}).
\end{equation}
While this expansion is convenient for obtaining analytical expressions for matrix elements, it involves five angular variables, and is thus not well suited for computational purposes. A convenient way of evaluating the $V_{\lambda\mu}$ coefficients is outlined in Appendix C.

Making use of standard expressions for the integrals involving products of three $D$-functions and spherical harmonics \cite{Zare}, we obtain the final result for the matrix element of the interaction potential 
\begin{multline}\label{me_V}
\langle N M_N K_N | \langle S M_S | \langle l m_l | V(\bm{R},\hat{\Omega}) | N' M_N' K_N' \rangle | S M_S'\rangle | l' m_l' \rangle = \\
\delta_{M_S,M_S'} [(2N+1)(2N'+1)(2l+1)(2l'+1)]^{1/2} \times (-1)^{M_N'-K_N'+m_l} \\
\times \sum_{\lambda= 0}^{\lambda_\text{max}} \sum_{\mu \ge 0}^\lambda V_{\lambda\mu} (R) \frac{1}{1+\delta_{\mu 0}} 
\sum_{m_\lambda} \threejm{N}{M_N}{\lambda}{m_\lambda}{N'}{-M_N'} \threejm{l}{-m_l}{\lambda}{m_\lambda}{l'}{m_l'}  \threejm{l}{0}{\lambda}{0}{l'}{0} \\
\biggl{[} \threejm{N}{K_N}{\lambda}{\mu}{N'}{-K_N'} + (-1)^\mu \threejm{N}{K_N}{\lambda}{-\mu}{N'}{-K_N'} \biggr{]}.
 \end{multline}
Setting $K_N=0$ reduces this expression to the familiar Legendre expansion of the atom-diatomic molecule interaction potential \cite{Lester,Roman04}. As in the  atom-diatom case \cite{Roman04}, the interaction potential $V(\bm{R},\hat{\Omega})$ does not couple basis functions with different $M_S$. 

The coefficients $V_{\lambda\mu}(R)$ were evaluated as described in Appendix C using a  $15\times 15$ direct-product Gauss-Legendre quadrature grid in $\theta$ and $\phi$. The expansion (\ref{me_V}) was truncated at $\lambda_\text{max}=7$ and we verified that increasing $\lambda_\text{max}$ does not affect the numerical results. The system of CC equations (\ref{CC}) was solved numerically using the log-derivative algorithm \cite{David} on a grid of $R$ extending from $R_\text{min}=3.0a_0$ to $R_\text{max}=40.04 a_0$ with a grid step of $0.04a_0$ for collision energies larger than 0.1 cm$^{-1}$. For ultralow collision energies ($E_C<0.1$ cm$^{-1}$), $R_\text{max}$ was increased to $80.04a_0$ (for $10^{-4}$ cm$^{-1}<E_C<0.1$ cm$^{-1}$) and 160.04$a_0$ ($10^{-6}$ cm$^{-1}<E_C<10^{-4}$ cm$^{-1}$).

To obtain the cross sections converged to within $<$10\%, the coupled-channel basis set included  $5$ rotational states of CH$_2$ augmented with 5 (for $E_C<1$ cm$^{-1}$) or 6 (for $E_C>1$ cm$^{-1}$) partial waves. The same basis set parameters were used for all CH$_2$ isotopomers.


\section{Results}

\subsection{Ab initio PESs}

In this work, we use two different {\it ab initio} potential energy surfaces (PESs) for the He-CH$_2$ van der Waals complex: PES A and PES B. The former (PES A) was utilized in our previous theoretical work on He + CH$_2$  scattering in a magnetic field \cite{prl11}. A more recent, high-quality PES B developed by Ma {\it et al.} \cite{MillardCH2new}  has  been used in quantum calculations of rotational relaxation in He + CH$_2$ collisions at room temperature.  

PES A was calculated using a UCCSD(T) method (coupled-cluster including single and double excitations along with a perturbative correction for  triple excitations) \cite{WAT93:8718} and a large cc-pVTZ  basis set \cite{WOO93:1358}.  For fitting purposes, the PES is described by an analytic function that sums over six pairs of interaction potentials,\cite{TAN84:3726}
\begin{equation}
\label{eq:pes}
V_{{\rm He-CH}_2}=\sum_{\alpha} \biggl\{ A_{\alpha}e^{-b_{\alpha}R_{\alpha}}
-f_6(b_{\alpha}R_{\alpha})\frac{C_{6,\alpha}}{R_{\alpha}^6} \biggr\},
\end{equation}
with the damping function
\begin{equation}
f_n(x)=1-e^{-x}\sum_{k=0}^{n}\frac{x^k}{k!}.
\end{equation}
The six pairs refer to $\alpha=\{ {\rm He-C}, {\rm He-H}_a, {\rm He-H}_b, {\rm He-G}_1, {\rm He-G}_2,  {\rm He-G}_3 \}$, where G$_i$ are three pseudo-atoms. Their MF coordinates relative to C are fixed at (0,1,0)\,$a_0$ for G$_1$, (0,-1,0)\,$a_0$ for G$_2$, and (0,0,1)\,$a_0$ for G$_3$. The pseudo-atoms are used to take the many-body interaction potentials into account.

 The parameters in Eq.~(\ref{eq:pes}) are determined by fitting to 985 CCSD(T)/cc-pVTZ points spanning the range of $R=[3.0, 7.5]\,a_0$, $\theta=[3^0, 179^0]$, and $\phi=[5^0,175^0]$ with a fixed geometry of CH$_2$ at its triplet equilibrium structure. The potential energy surface is well fitted to a root-mean-square error of 0.8 cm$^{-1}$. The parameters obtained are given in Table~\ref{tab:pes}. 

\begin{figure}[t]
	\centering
	\includegraphics[width=1.05\textwidth, trim = 0 0 0 0]{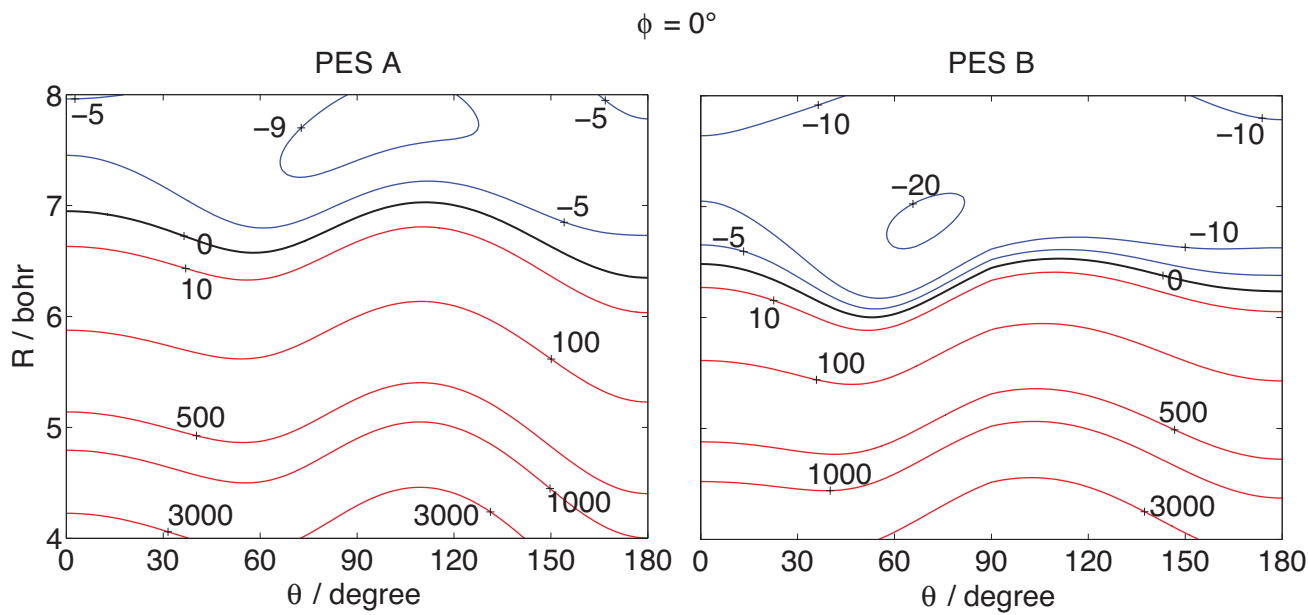}
	\renewcommand{\figurename}{Fig.}
	\caption{Contour plots of PES A (left) and PES B (right) for $\phi = 0^\circ$. $\theta$ and $\phi$ are the spherical polar angles of vector $\mathbf{R}$ in the MF frame (see Fig. \ref{fig:MF} and text for details.) Energies are in cm$^{-1}$.}
	\label{fig:pesBA1}
\end{figure}

\begin{figure}[t]
	\centering
	\includegraphics[width=1.05\textwidth, trim = 0 0 0 0]{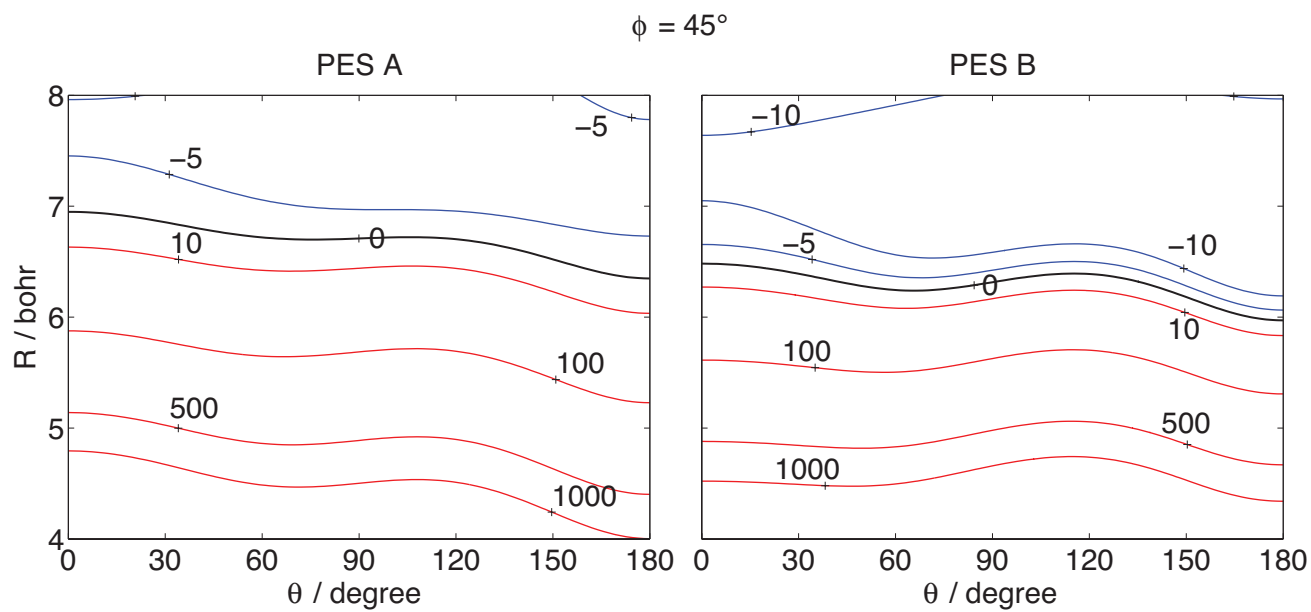}
	\renewcommand{\figurename}{Fig.}
	\caption{Same as in Fig. \ref{fig:pesBA1} for $\phi = 45^\circ$.}
	\label{fig:pesBA2}
\end{figure}

\begin{figure}[t]
	\centering
	\includegraphics[width=1\textwidth, trim = 0 0 0 0]{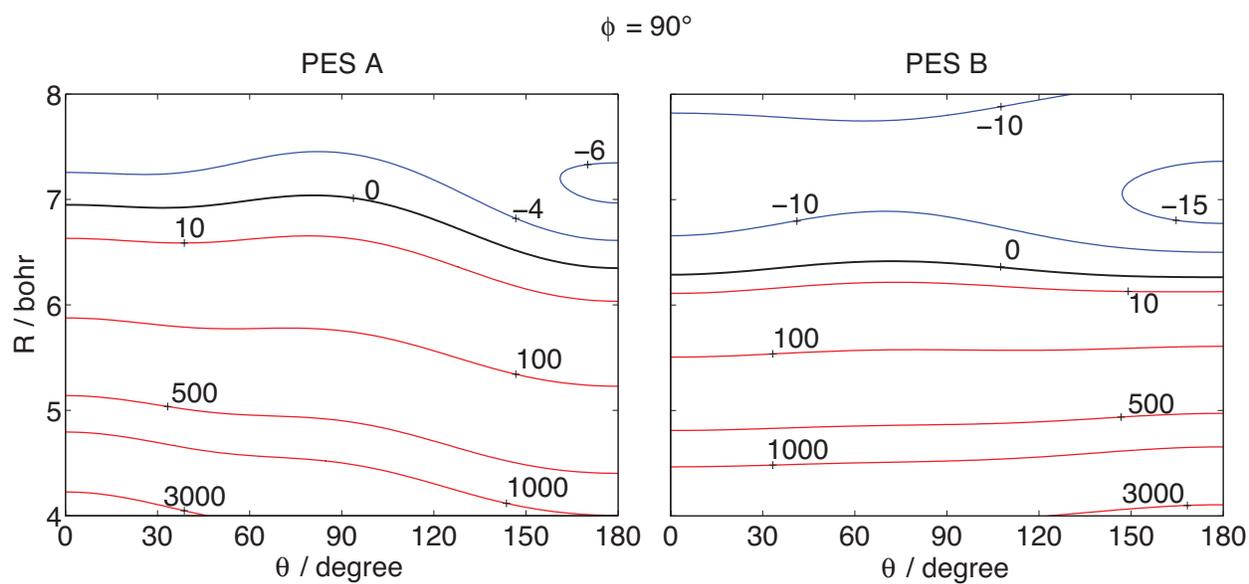}
	\renewcommand{\figurename}{Fig.}
	\caption{Same as in Fig. \ref{fig:pesBA1} for $\phi = 90^\circ$.}
	\label{fig:pesBA3}
\end{figure}

Figures~\ref{fig:pesBA1} -- \ref{fig:pesBA3} (left panels) show a typical long-range potential energy surface for the He$-$CH$_2$ complex, where the CH$_2$ molecule is fixed at its CCSD(T) optimized  geometries, $R_{\rm CH}=2.0404\,a_0$ and $\theta_{\rm HCH}=133.11^0$. It displays two equivalent global minima located at $R_{{\rm He-CH}_2}=7.68\,a_0$ with a well depth of 9.7\,cm$^{-1}$. Owing to the shallow minima, the interaction potentials between the two collision partners are weakly anisotropic. This feature of the potential energy surface is also consistent with the small permanent electric dipole moment of 0.2293 atomic units ($0.5828$\,D) calculated with CCSD(T)/cc-pVTZ and nearly isotropic quadrupole moments [(6.0344, 4.3788, 5.6653) atomic units] of CH$_2$.

A detailed description of the {\it ab initio} approach and the topology of PES B has been presented elsewhere \cite{MillardCH2new}. Spin-restricted coupled-cluster [RCCSD(T)] calculations  \cite{L1,L2} were carried out on a 4-dimensional grid consisting of the position (in spherical polar coordinates) of the He atom with respect to the center-or-mass of the CH$_2(\tilde{X})$ molecule for a grid of values of the CH$_2$ bond angle.  An aug-cc-pvqz \cite{L3} basis set was used with a set of bond functions \cite{L4,L5} added at the mid-point of $R$, the vector pointing to the He atom from the center of mass of CH$_2$. The 3-dimensional PES used in the scattering calculations was determined by averaging the 4-dimensional potential over the bending angle weighted by the square of the $v_b=0$  bending vibrational wave function.

Figures~\ref{fig:pesBA1} -- \ref{fig:pesBA3} show that PES A is considerably less deep, more repulsive at short range, and less anisotropic than PES B.  The increased depth of the potential arises from a better recovery of the correlation energy, which is known to be responsible for dispersion interactions, due to the inclusion of diffuse atom-centered functions as well as bond functions \cite{Chalasinski} for a weakly-bound complex. The calculations on which PES A is based are deficient in that the van der Waals attraction is significantly less well described. We also note that because the van der Waals well is deeper in PES B, the repulsive core of the PES -- the angle-dependent hard sphere radius -- is significantly smaller in extent than the predictions of PES A.
As will be shown below (Sec. IIIB),  the significant difference in the two PESs has a strong effect on both the elastic and inelastic He + CH$_2(\tilde{X})$ collisions at low temperatures.




\subsection{Low-temperature He + $o$-CH$_2$ collisions in a magnetic field}

\begin{figure}[t]
	\centering
	\includegraphics[width=0.85\textwidth, trim = 0 0 0 0]{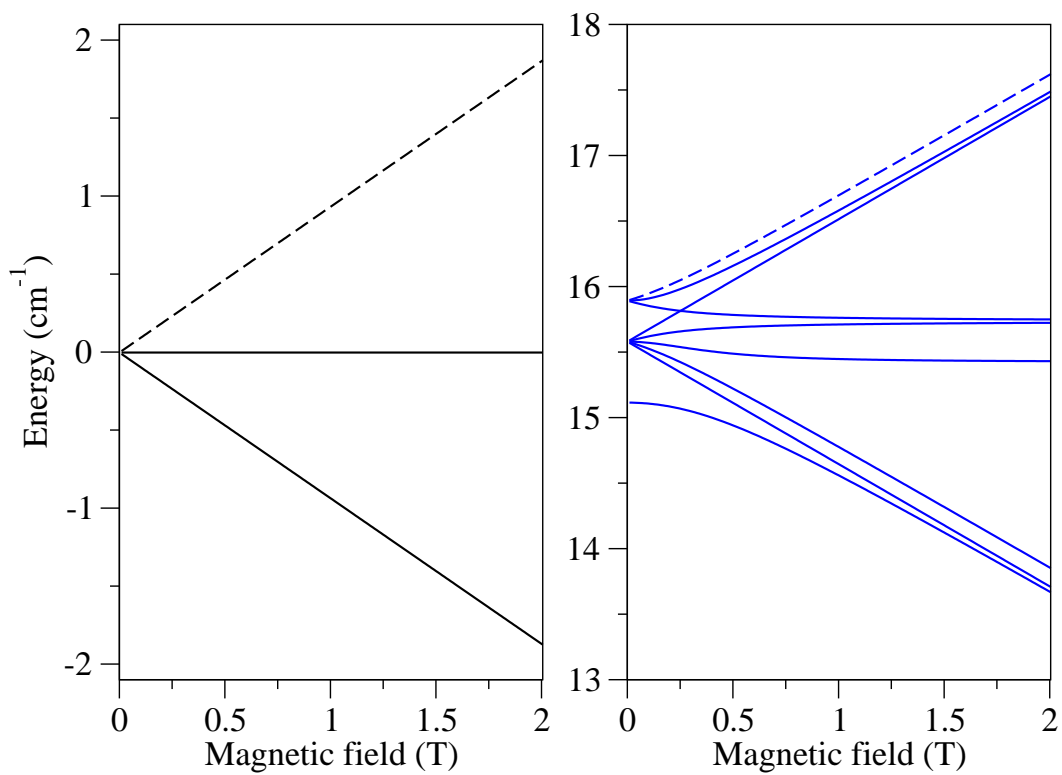}
	\renewcommand{\figurename}{Fig.}
	\caption{Lowest Zeeman energy levels of $o$-CH$_2$ (left panel) and $p$-CH$_2$ (right panel). The initial states used in scattering calculations are indicated by dashed lines.}\label{fig:Zeeman}
\end{figure}

Figure \ref{fig:Zeeman} shows the Zeeman energy levels of {\it ortho} and {\it para} nuclear spin isomers of CH$_2$. The ground  rotational state of $o$-CH$_2$ ($0_{00}$)  is split by magnetic fields into three Zeeman sublevels with $M_S=-1$, 0, and 1.  Here, we adopt the conventional $N_{K_{o} K_{p}}$  notation for the rotational states of an asymmetric top,  where $K_{o}$ ($K_{p}$) are the projections of $\hat{\bm{N}}$ on the MF quantization axis in the oblate (or prolate) symmetric top limits \cite{ZareNote}. The ground rotational state of $p$-CH$_2$ ($1_{01}$) has $N=1$ \cite{note_rot,Bunker} and splits into nine Zeeman sublevels correlating with the $j=0$, $j=1$, and $j=2$ fine-structure levels in the limit of zero magnetic field, where $j=N+S$ is the total angular momentum of CH$_2$ excluding nuclear spin. Here, we are interested in collisions of CH$_2$ molecules in their magnetically trappable, maximally spin-stretched Zeeman levels $M_S=+1$ (for $o$-CH$_2$) and $m_j = +1$ (for $p$-CH$_2$), where $m_j = M_S+M_N$ is the projection of $j$ on the magnetic field axis. It is clear from Fig. \ref{fig:Zeeman} that the Zeeman energy level patterns of $o$-CH$_2$ and $p$-CH$_2$ are very different. As will be shown below, this difference leads to dramatic variations in the low-temperature collisional properties of the spin isomers. In this section, we focus on collisions of $o$-CH$_2$ molecules and omit the prefix $o$  for brevity.

\begin{figure}[t]
	\centering
	\includegraphics[width=0.85\textwidth, trim = 0 0 0 0]{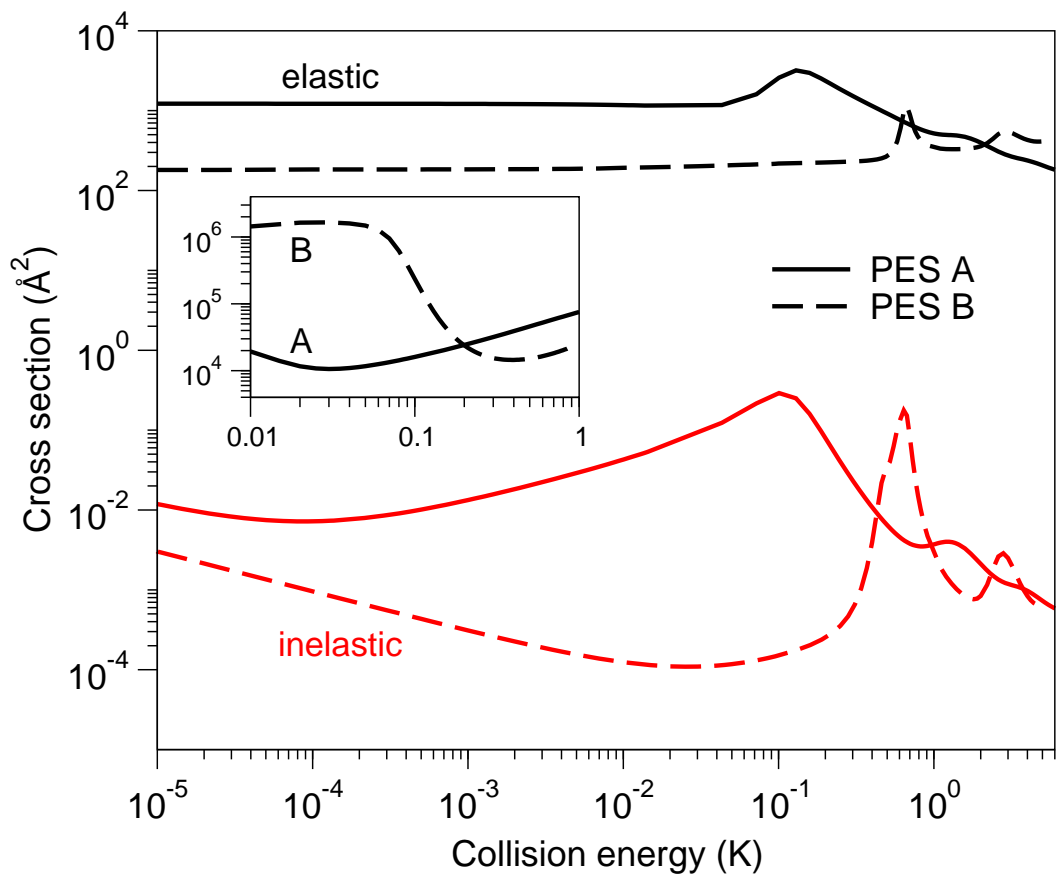}
	\renewcommand{\figurename}{Fig.}
	\caption{Collision energy dependence of elastic and inelastic cross sections for He-CH$_2$ calculated using PES A (full lines) and PES B (dashed lines). The magnetic field is 0.1 T. The inset shows the thermally averaged ratio of elastic to inelastic collision rates as a function of temperature.}\label{xs_potAB}
\end{figure}

Figure \ref{xs_potAB} shows the elastic and inelastic cross sections for $^3$He + CH$_2(0_{00},\,M_S=+1)$ collisions \cite{note_He_isotope} as functions of collision energy calculated for $B=0.1$ T, the field strength typically used in  buffer-gas cooling and magnetic trapping experiments with $^3\Sigma$ molecules \cite{HeNH07,HeNH}. 
  We calculate the inelastic spin relaxation cross section by summing the state-to-state cross sections for a given incident collision channel (dashed lines in Fig. \ref{fig:Zeeman}) over all energetically accessible Zeeman levels of CH$_2$ (full lines in Fig. \ref{fig:Zeeman}). Both the elastic and inelastic cross sections are very sensitive to the interaction PESs: the $s$-wave limit of the elastic cross section calculated using PES A is $\sim$10 times larger than the limit for PES B. Overall, because calculations on which PES A is based predict a larger hard-sphere radius, we anticipate that the $s$-wave elastic cross section, which is a measure of the size of the system,  will be larger. 

The cross sections calculated using PES A are dominated by a pronounced resonance feature at $E_C\sim 0.1$ K with a very broad shoulder extending down to $E_C\sim 10^{-4}$ K, whereas the results for PES B display two resonance peaks above $E_C=0.1$~K. The presence of the broad resonance leads to a large difference between the thermally averaged ratios of elastic to inelastic collision rates calculated with potentials A and B (see the inset of Fig. \ref{xs_potAB}).  Since PES B has a more pronounced long-range attractive component, we anticipate that any elastic resonances at low energy predicted by the two PES's will be significantly different.

\begin{figure}[t]
	\centering
	\includegraphics[width=0.85\textwidth, trim = 0 0 0 0]{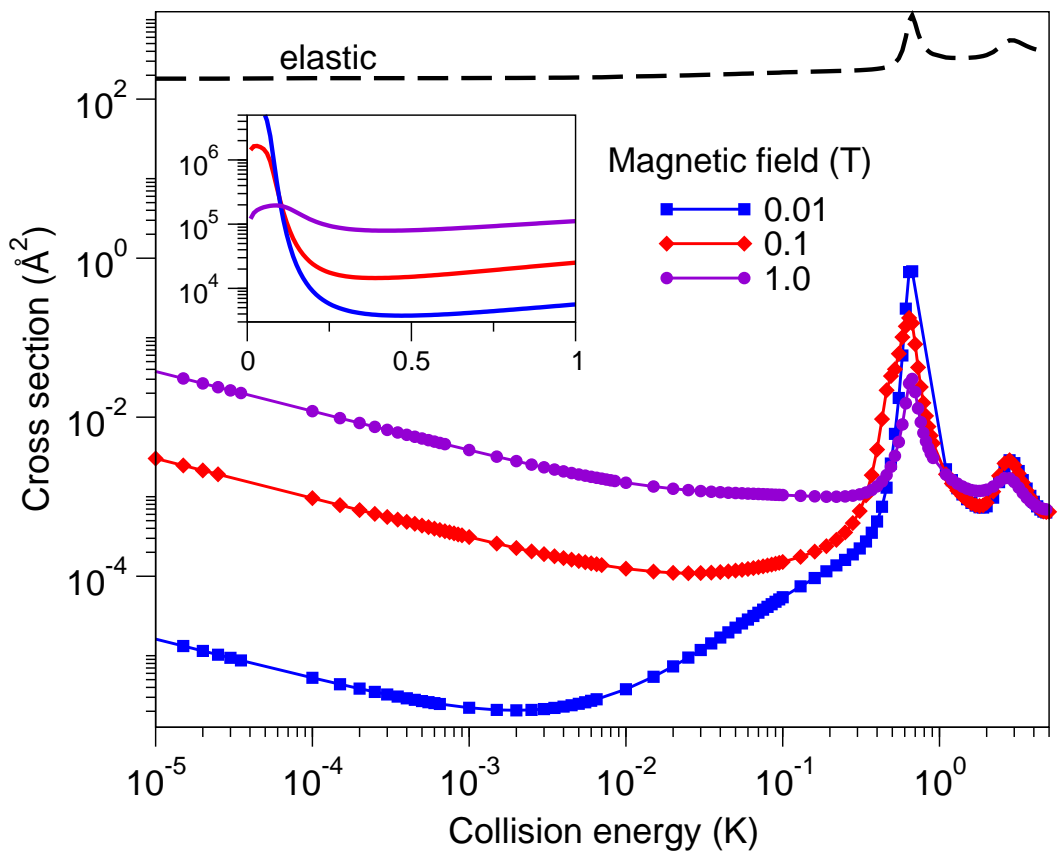}
	\renewcommand{\figurename}{Fig.}
	\caption{Collision energy dependence of the inelastic cross sections for He-CH$_2$ calculated  for several values of the magnetic field: 0.01 T (squares), 0.1 T (diamonds), and 1 T (circles). The inset shows the temperature dependence of the ratio of elastic to inelastic collision rates for the same values of the magnetic field. All the results are obtained using PES B.}\label{xs_B}
\end{figure}

Figure \ref{xs_B}  shows the cross sections for inelastic spin relaxation in He + CH$_2$ collisions as functions of collision energy calculated for several values of the magnetic field. At very low collision energies, we observe the  $s$-wave threshold behavior $\sigma_\text{inel}\sim 1/\sqrt{E}_C$ typical of two-body inelastic processes in the limit of zero collision energy \cite{Wigner}. The elastic cross section is independent of the field and becomes a constant at $E_C<$ 100 mK. At moderately low collision energies (above 0.5 K), the cross sections exhibit resonant oscillations due to the excited rotational states ($l>0$) of the collision complex (shape resonances) which decay by tunneling through the centrifugal barrier \cite{HeNH,Zygelman}. A crossover between the single $s$-wave and multiple partial-wave regimes occurs at a collision energy of 1 mK at low magnetic fields ($B=0.01$ T) and shifts to higher collision energies with increasing field. A similar trend manifests itself in cold collisions of $^3\Sigma$ molecules \cite{Roman04}. Figure \ref{xs_B} shows that the magnitude of the inelastic cross section in the $s$-wave limit increases dramatically with the field strength. As in previous theoretical work on  collisions of  $\Sigma$-state molecules, this can be explained by the following argument \cite{VolpiBohn,Roman04}. The height of the centrifugal barrier in the outgoing ($M_S=0,-1$) collision channels becomes smaller with increasing field, making it easier for collision products to escape over the barrier, thereby enhancing the inelastic collision rate \cite{VolpiBohn, Roman04,Roman08}.

In the inset of Fig. \ref{xs_B}, we plot the key figure of merit for sympathetic cooling experiments -- the ratio of the thermal rate constants for elastic scattering and inelastic scattering (spin relaxation) in He-CH$_2$ collisions:
\begin{equation}\label{gamma}
\gamma (T) = \frac{k_\text{el}(T)}{k_\text{inel}(T)}.
\end{equation}
The rates in Eq. (\ref{gamma}) are obtained as functions of temperature by averaging the cross sections shown in Fig. \ref{xs_B} over a Maxwell-Boltzmann velocity distribution. 
We observe that $\gamma>1000$  over the range of magnetic fields typically used in cryogenic cooling and magnetic trapping experiments (0.01 -- 1 T).  An empirical rule states that $\gamma>100$ is required  to ensure efficient cryogenic He buffer-gas cooling  of a molecule \cite{BufferGasCooling}, so the results shown in Fig.~\ref{xs_B} indicate that $p$-CH$_2$ is an ideal candidate for cryogenic cooling and magnetic trapping experiments \cite{BufferGasCooling}. As shown in the inset of Fig. \ref{xs_potAB}, the temperature profiles of $\gamma$ calculated for PESs A and B differ  substantially, with  $\gamma_\text{A} \gg \gamma_\text{B}$ at $T<0.1$ K due to the broad scattering resonance that occurs on PES A (see the inset of Fig.~\ref{xs_potAB}).  We emphasize, however, that despite these quantitative differences, both PES A and PES B calculations predict $\gamma(T)>100$, so our main qualitative conclusion as to the suitability of CH$_2$ for sympathetic cooling experiments is independent of the choice of the PES.



\subsection{Spin relaxation of nuclear spin isomers}

So far we have been focusing on collisions involving a single nuclear spin isomer of CH$_2$ ({\it ortho}-CH$_2$). However, as pointed out in Sec. II, the rotational and Zeeman structures of $o$- and $p$-CH$_2$ are very different, so it is of interest to compare the collisional properties of the nuclear spin isomers. Figure \ref{xs_op} shows the elastic and inelastic cross sections for $o$-CH$_2$ and $p$-CH$_2$ in He as functions of collision energy.   The inelastic cross section calculated for $p$-CH$_2$ is about four orders of magnitude larger than that for $o$-CH$_2$ over the range of collision energies from 10$^{-5}$ to $0.1$ K.  In the multiple partial wave regime ($E_C>0.1$ K) the difference becomes somewhat smaller, but never falls below three orders of magnitude.

In order to understand this remarkable disparity in low-temperature collisional properties of different nuclear spin isomers of CH$_2$,  we note that in our calculations, the incident collision channels for both $ortho$ and $para$-CH$_2$ are the maximally spin-stretched Zeeman states (shown in Fig. \ref{fig:Zeeman} by dashed lines). These are the Zeeman states in which the molecules reside when confined in a permanent magnetic trap \cite{Roman08,NJP,BufferGasCooling}.  However, the $rotational$ manifolds to which these Zeeman states belong are different for $ortho$ and $para$ nuclear spin isomers:  As shown in the left panel of Fig. \ref{fig:Zeeman}, the lowest maximally spin-stretched Zeeman state of $o$-CH$_2$ belongs to the $0_{00}$ rotational manifold, while that for $p$-CH$_2$ belongs to the $1_{01}$ rotational manifold \cite{note_rot}.

The spin-spin interaction (\ref{Hss_BF}) couples the ground $0_{00}$ and the second excited $2_{02}$ rotational manifolds of $o$-CH$_2$. As a result, both $N$ and $M_S$ lose the status of good quantum numbers, and the wavefunction of the lowest magnetically trappable Zeeman state becomes
\begin{equation}\label{initial_state_o}
|\tilde{0}_{00}, \tilde{M}_S= 1\rangle = \alpha_0 |0_{00}, M_S = 1\rangle +  \beta_0 |2_{02}, M_S=0\rangle + \gamma_0 |2_{02}, M_S=-1\rangle,
\end{equation}
where the tildes over $N$ and $M_S$ indicate the approximate nature of these quantum numbers. A first-order perturbation theory estimate gives 
\begin{equation}\label{beta0}
\beta_0 = - \frac{ \langle 0_{00}, M_N=0| \langle 11 |\hat{H}_\text{ss}| 2_{02}, M_N' \rangle | 10\rangle}{\Delta E_{02}} \sim \frac{D}{\Delta E_{02}},
\end{equation}
where the basis functions are the products of asymmetric-top eigenfunctions $|N_{K_o K_p}M_N\rangle $ times the spin functions $|SM_S\rangle$, and we have neglected the dependence of the matrix element of the spin-spin interaction (\ref{Hss_BF}) on the zero-field-splitting parameter $E$ assuming $E \ll D$, which is justified for CH$_2(\tilde{X})$ where $E/D = 0.051$. In Eq. (\ref{beta0}) \cite{ZareNote}
\begin{equation}\label{DeltaE02}
\Delta E_{02} = 2(A+B+C) -2[(B-C)^2 + (A-C)(A-B)]^{1/2}
\end{equation}
is the energy gap between the ground and the second excited rotational states. In $o$-CH$_2$ the gap (47 cm$^{-1}$) is much larger than the spin-spin coupling $D=0.78$ cm$^{-1}$, so the admixture of the $2_{02}$ rotational state into the ground state is on the order of a few percent (depending on $M_N'$).

In the first Born approximation, the cross section for spin relaxation out of the lowest magnetically trappable state of CH$_2$ ($|\tilde{0}_{00}, \tilde{M}_S= 1\rangle$) is proportional to the square of the matrix element between the molecular spin states involved in the transition. In particular, the cross section for relaxation  from the $\tilde{M}_S=+1$ state in the lowest rotational manifold of $o$-CH$_2$ may be written as \cite{VolpiBohn} $\sigma_{\tilde{M}_S \to \tilde{M}_S'} \sim |\langle \tilde{0}_{00}, \tilde{M}_S | V(R,\theta,\phi) | \tilde{0}_{00}, \tilde{M}_S'\rangle|^2$. 
Using the basis functions given by Eq. (\ref{initial_state_o}), we find for the $\tilde{M}_S=1\to M_S'=-1$ transition to the leading order
\begin{equation}\label{Born2}
\sigma_{1 \to 0} \sim \beta_0^2 \times | \langle 2_{02} |V(R, \theta,\phi)| 0_{00}\rangle|^2 \sim  \left( \frac{D}{\Delta E_{02}}\right)^2 | \langle 2_{02} |V(R, \theta,\phi)| 0_{00}\rangle|^2.
\end{equation}
This result shows that spin relaxation in spin-1 molecular radicals is a first-order process mediated by the combined action of the spin-spin interaction and the anisotropy of the interaction potential (the second term in Eq. \ref{Born2}). Note that both these interactions are equally important: without the spin-spin interaction there would be no admixture of the spin-down state in the initial spin-up state ($D=0$ implies $\beta_0=0$ in Eq. \ref{beta0}), and without the potential anisotropy there would be no coupling between the different $N$ states.

 We further note that the expression (\ref{Born2}) contains the factor $D^2/\Delta E_{02}^2\ll 1$ for light molecules with large rotational splittings (\ref{DeltaE02}). Thus, collision-induced spin relaxation in such molecules should be  slower than rotational relaxation, which is induced by the anisotropic part of the interaction potential {\it without} the small factor $D^2/\Delta E_{02}^2$. These considerations explain the small magnitude of the calculated inelastic cross sections for $o$-CH$_2$ shown in Fig. \ref{xs_B}.  A closely related result for collision-induced spin relaxation in linear $^3\Sigma$ molecules was first derived theoretically by Krems and co-workers \cite{Roman04} and then observed experimentally using He buffer-gas cooled  NH($^3\Sigma$) molecules in a  magnetic trap at 600~mK \cite{HeNH}.






The wavefunction for the lowest magnetically trappable Zeeman state of $p$-CH$_2$ may be written as
\begin{equation}\label{initial_state_p}
|\tilde{1}_{01}, \tilde{M}_S= 1\rangle = \alpha |1_{01}, M_S = 1\rangle +  \beta_1 |1_{01}, M_S=0\rangle + \gamma_1 |1_{01}, M_S=-1\rangle.
\end{equation}
Unlike the ground rotational state of $o$-CH$_2$  the spin-rotation and spin-spin interactions couple the different $M_N$ and $M_S$ sublevels {\it within} the $1_{01}$ rotational manifold of $p$-CH$_2$. Thus, the mixing coefficients $\alpha_1,\beta_1$, and $\gamma_1$ in Eq. (\ref{initial_state_p}) are all of similar magnitude, and  the cross section for spin relaxation in $p$-CH$_2$ collisions is given by an expression similar to Eq.~(\ref{Born2}) {\it without} the small factor $D^2/\Delta E_{02}^2$ (assuming the validity of Born approximation). 
Thus the mechanism of spin relaxation in He + $p$-CH$_2$ collisions is similar to that of rotational relaxation or rotational depolarization with the corresponding cross sections ranging from  1 to 100  \AA$^2$ (Ref. \cite{MillardCH2new}). This lack of suppression is responsible for the large magnitude of spin relaxation cross sections for He + $p$-CH$_2$  shown in Fig. \ref{xs_op},  explaining the large differences between the collisional properties of different nuclear spin isomers of CH$_2$ in low-temperature collisions with He atoms.

A simple qualitative explanation for this difference is as follows:  Inelastic changes in the spin-multiplets correspond to a reorientation of $\vec N$ followed by a recoupling with $\vec S$, which is unaffected by the intermolecular potential which has no magnetic terms.  In the case of the lowest low-field-seeking Zeeman state of $o$-CH$_2$ ($|\bar{0}_{00}, \bar{M}_S=+1\rangle$), the nonzero $\vec{N}$ component is only a small admixture in the ground-state wavefunction (see Eq. \ref{initial_state_o}) so reorientation occurs at a much slower rate.


\subsection{Isotopic effects}

In the previous section, we have established that the cross section of spin relaxation in ground-state $o$-CH$_2$ collisions scales as $D^2/\Delta {E}^2_{02}$ with the splitting $\Delta E_{02}$ between the ground $0_{00}$ and the second excited $2_{02}$ rotational levels  of $o$-CH$_2$. This scaling law is based on a simple first Born approximation, which does not take into account the effects of higher-order potential couplings and  shape resonances (see Fig. \ref{xs_op}) on collision dynamics. It is therefore desirable to test the accuracy of the scaling prediction, which can be performed by changing the rotational constant (and hence the energy gap $\Delta E_{02}$) of CH$_2$. To this aim, we have performed quantum  scattering calculations for partially and fully deuterated isotopomers of CH$_2$ using the same PES B for all the isotopomers. The rotational and zero-field-splitting constants of CH$_2$, CHD, and CD$_2$  used in these calculations are listed in  Table II \cite{cmshift}.

\begin{figure}[t]
	\centering
	\includegraphics[width=0.85\textwidth, trim = 0 0 0 0]{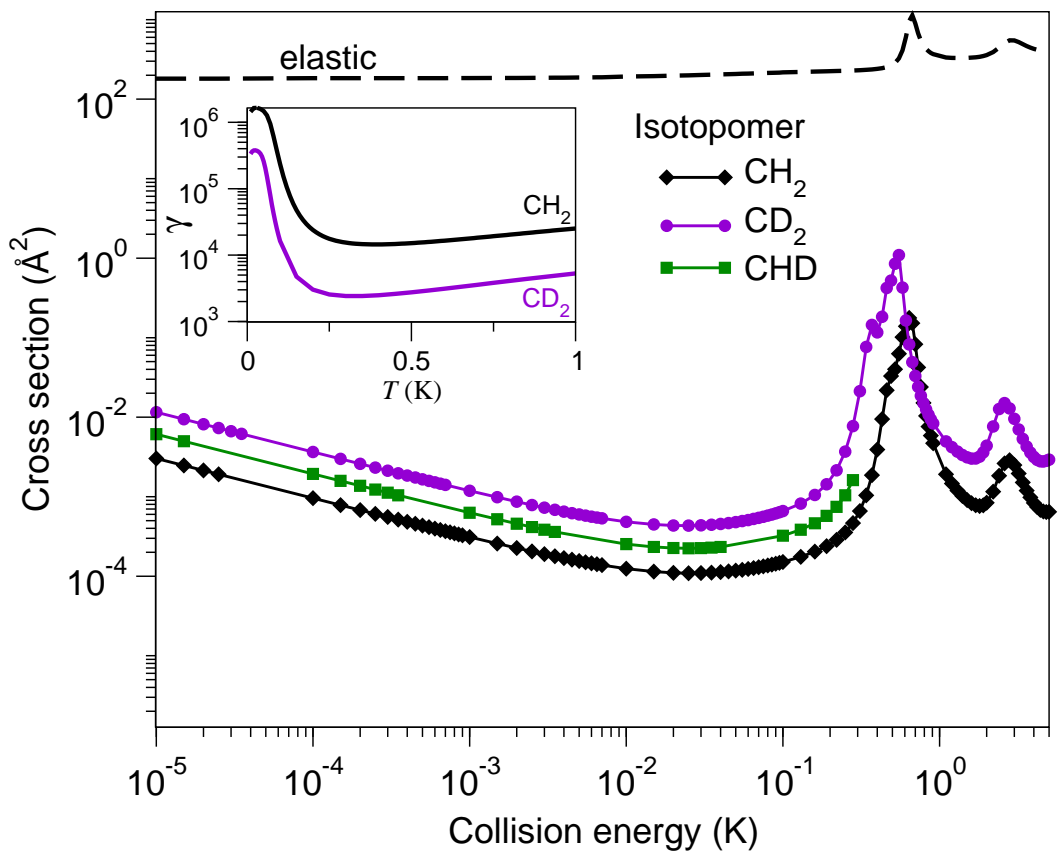}
	\renewcommand{\figurename}{Fig.}
	\caption{Collision energy dependence of the inelastic cross sections for collisions of different methylene isotopomers with $^3$He:  $o$-CH$_2$ (diamonds), CD$_2$ (circles) and CHD (squares). The magnetic field is 0.1 T. The inset shows the temperature dependence of the ratio of elastic to inelastic collision rates for the different isotopomers. All the results are obtained using PES B.}\label{xs_isotopes}
\end{figure}

Figure \ref{xs_isotopes} compares the cross sections for spin relaxation in CH$_2$, CHD, and CD$_2$ induced by collisions with He at $B=0.1$ T. Remarkably, the cross sections exhibit a nearly identical dependence on collision energy in the few-partial-wave  regime ($E_C<100$ mK).  This result points to the validity of the inverse quadratic scaling of spin relaxation cross sections with the rotational splitting (\ref{Born2}). Table III lists the ratios of the cross sections for CHD and CD$_2$ to those of CH$_2$ calculated at two different collision energies. The ratios are seen to increase with a decreasing rotational splitting $\Delta E_{02}$. The deviation from the $D^2/\Delta E^2_{02}$ scaling does not exceed 10\%, demonstrating the araccuracy of the scaling prediction (\ref{Born2}) which is quite remarkable given its simplicity. 

In the multiple partial wave regime, the inelastic cross sections  in Fig. \ref{xs_isotopes} tend to be dominated by shape resonances, which occur due to the centrifugal barriers in the incoming and outgoing collision channels \cite{VolpiBohn,Zygelman}. These resonances lead to an enhancement of scattering cross sections that is correlated with the magnitude of the rotational constant of the metastable collision complex. The properties of this collision complex are determined not only by the rotational structure of the molecule, but also by the reduced mass and interaction potential of the He-molecule complex, leading to pronounced deviations from the simple $D^2/\Delta {E}^2_{02}$ scaling near scattering resonances. This effect is clearly observed in Fig. \ref{xs_isotopes}: In contrast with the scaling prediction, the inelastic cross section for CH$_2$ can {\it exceed} that of CD$_2$ at selected collision energies. We note that similar deviations from the $D^2/\Delta {E}^2_{02}$ scaling were observed experimentally for $^3$He + NH (but not $^4$He - NH) collisions at 600 K \cite{HeNH}, demonstrating the important role of shape resonances in atom-molecule collisions at sub-Kelvin temperatures.





\section{Summary and conclusions}

We have developed a theoretical methodology for exact quantum mechanical calculations of polyatomic molecule collisions  in the presence of an external magnetic field. The theoretical description is based on a fully uncoupled SF basis set expansion to represent the scattering  wave function. A system of coupled-channel equations for the radial part of the wave function is derived and parametrized by the matrix elements of the molecular Hamiltonian and of the atom-molecule interaction PES. Using CH$_2(\tilde X)$ as a general example of a non-linear, asymmetric-top polyatomic molecule,  we present the expressions for the matrix elements of rotational, centrifugal distortion, spin-rotation, and spin-spin interactions in terms of the relevant spectroscopic constants \cite{Bunker} (Sec. II and Appendices A and B). The matrix elements of the atom-molecule interaction are evaluated by expanding the PES  in angular basis functions, and we outline the details of this procedure  in Sec. II and Appendix~C. 

We have applied our theoretical framework to explore the dynamics of spin relaxation of ground-state CH$_2(\tilde X)$ molecules induced by collisions with He atoms in a magnetic field. The scattering calculations were based on two different {\it ab initio} PESs developed independently of each other: PES A \cite{prl11} and PES B  \cite{MillardCH2new}.  
We found that, even though the results obtained for PES A and B exhibit large quantitative differences, the ratio of elastic to inelastic collision rates for He + $o$-CH$_2$ collisions greatly exceeds 100 over a wide range of temperatures and magnetic fields regardless of the PESs used  to describe the interaction of He with CH$_2$. The magnitude of this ratio is thus relatively insensitive to the details of the long-range part of the potential, which is described quite differently by the two PESs, as well as to the overall ``size'' of the potential. 
The large magnitude of the ratio strongly suggests the possibility of cryogenic buffer-gas cooling and magnetic trapping of $o$-CH$_2$.

\begin{figure}[t]
	\centering
	\includegraphics[width=0.85\textwidth, trim = 0 0 0 0]{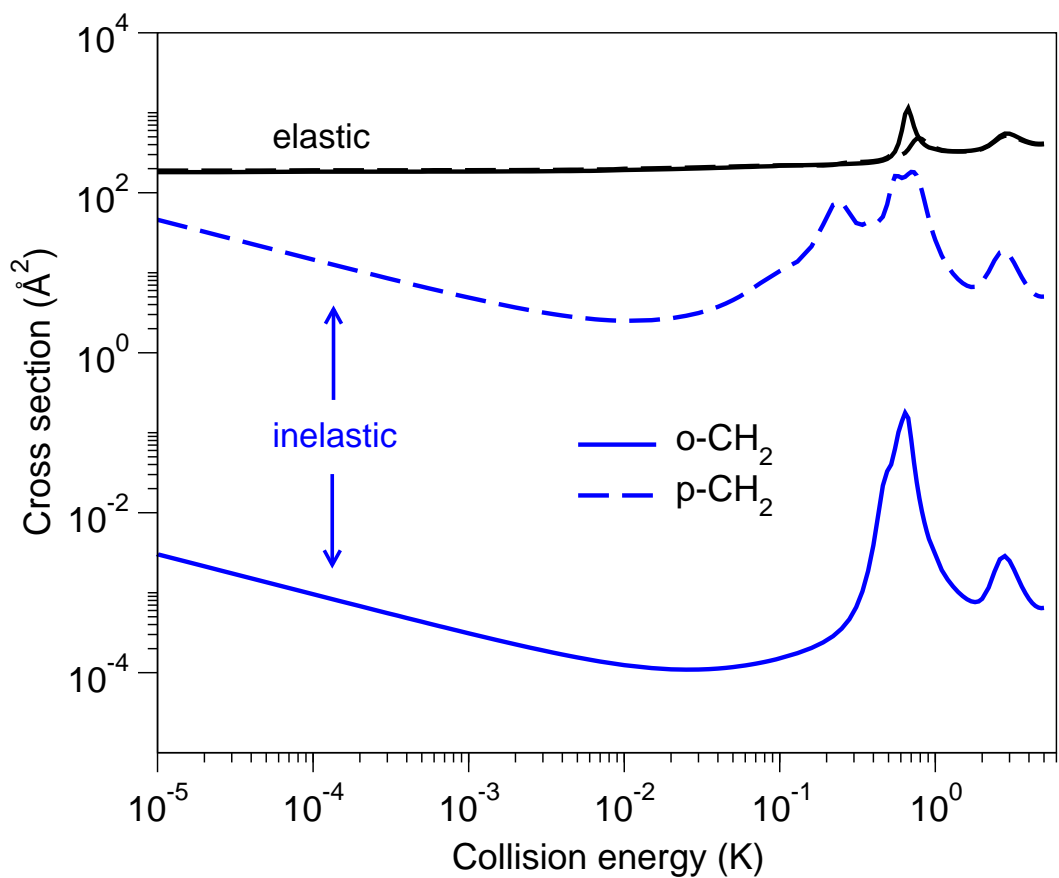}
	\renewcommand{\figurename}{Fig.}
	\caption{Collision energy dependence of the inelastic cross sections for $o$-CH$_2$ (full lines) and $p$-CH$_2$ (dashed lines) in He for $B=0.1$ T.  The results are obtained using PES B.}\label{xs_op}
\end{figure}

Our calculations demonstrate that different nuclear spin isomers of CH$_2$ undergo collision-induced spin relaxation on widely different timescales. As shown in Fig. \ref{xs_op}, the cross section for He-induced spin relaxation in $p$-CH$_2$ is more than three orders of magnitude larger than that for $o$-CH$_2$. These differences arise because the lowest $N=0$ rotational state, which is more immune to spin relaxation due to the absence of direct intramultiplet spin-rotation and spin-spin couplings,  is part of the rotational spectrum of $o$-CH$_2$, but not of $p$-CH$_2$.  The results shown in Fig. \ref{xs_op} thus suggest that  it may be possible to isolate the $ortho$ nuclear spin isomer of CH$_2$ by magnetic trapping of the naturally occurring $ortho$-$para$ mixture.  Separation of nuclear spin isomers is an active field of research \cite{separation}, and our results suggest that such separation will occur automatically when a cold $ortho$-$para$ mixture is confined in a magnetic trap in the presence of a cryogenic He gas. 





Using the first Born approximation, we have shown that  the cross sections for spin relaxation in the $N=0$ rotational state scale as   $D^2/\Delta E_{02}^2$ with the spin-spin coupling $D$ and the rotational  splitting $\Delta E_{02}$. This result is fully analogous to the previously derived scaling law for diatomic $^3\Sigma$ molecules \cite{Roman03,HeNH}. We verified that the deviation of the scaling predictions from exact CC calculations does not exceed 10\% in the $s$-wave regime (Fig. \ref{xs_isotopes} and Table III). The scaling prediction becomes less accurate with increasing collision energy, as the collision system enters the multiple partial wave regime (Fig. \ref{xs_isotopes}) and shape resonances occur.

The results of our analysis bear important implications for  collisional cooling of polyatomic molecules,  opening up the possibility of cryogenic buffer-gas cooling of CH$_2(\tilde X)$ and possibly other \cite{prl11} open-shell polyatomic molecules in a magnetic trap. Experimental realization of this possibility may open up new directions of research in the study of reaction intermediates \cite{CarbenesBook}, large molecule spectroscopy and photochemistry, and may even enable external field control of complex organic reactions \cite{Roman08}. 
Moreover, since the magnitudes of the elastic-to-inelastic ratios for He~+~CH$_2$ plotted in Fig. \ref{xs_op} are similar to those measured and calculated for He + NH($^3\Sigma$) \cite{HeNH,Roman03}, it is reasonable to assume that collisional properties of CH$_2$ molecules with atoms other than He will be similar to those of the isoelectronic NH$(^3\Sigma)$ molecule \cite{HeNH07,HeNH}. If so, it may be possible to  use sympathetic cooling with laser-cooled alkaline earth or spin-polarized nitrogen atoms \cite{MgNH,N-NH} to create  ultracold ensembles of CH$_2$, CHD, CD$_2$, and possibly other spin-1 and spin-1/2 polyatomic molecules \cite{prl11}. A more detailed theoretical study would be needed to fully explore this possibility, and could be done using the formalism presented in this work.

\acknowledgements

We thank Gregory E. Hall, John M. Doyle, and Paul J. Dagdigian for discussions. This work was supported by the DOE Office of Basic Energy Science and NSF grants to the Harvard-MIT CUA and ITAMP at Harvard University and the Smithsonian Astrophysical Observatory as well as to MHA at the University of Maryland (Grants CHE-0848110   and CHE-1213332).

\section*{Appendix A:  spin-spin interaction in SF frame}

Here, we present the derivation of Eq. (\ref{Hss_BF}) from the standard MF expression for the spin-spin interaction in polyatomic molecules \cite{Bunker,BunkerBook,Raynes}
\begin{align}\label{A1} 
H_\text{SS} =  {\textstyle \frac{D}{3}} ( 2\hat{S}_x^2 - \hat{S}_z^2 - \hat{S}_y^2 ) + E(\hat{S}_z^2 - \hat{S}_y^2),
\end{align}
where the subscripts $x,y$, and $z$  refer to MF Cartesian components of the spin operators. The MF and SF spherical tensor components of $\hat{\bm{S}}$ are related by \cite{Zare}
\begin{equation}\label{A2}
\hat{S}^{(1)}_{q}  = \sum_p \mathcal{D}^1_{p,q} (\hat{\Omega}) \hat{S}_{p}^{(1)}.
\end{equation}
Here and below, the index $q$ ($p$) is used to denote the MF (SF) components of $\hat{\bm{S}}$ \cite{Zare}. A product of two MF spherical tensor operators can thus be expanded as 
\begin{equation}\label{A3}
\hat{S}^{(1)}_{q_1} \hat{S}^{(1)}_{q_2} = \sum_{p_1,p_2} \sum_k \cgc{1}{q_1}{1}{q_2}{k}{q_1+q_2} \cgc{1}{p_1}{1}{p_2}{k}{p_1+p_2} \mathcal{D}^k_{p_1+p_2,q_1+q_2} (\hat{\Omega}) \hat{S}_{p_1}^{(1)} \hat{S}_{p_2}^{(1)},
\end{equation}
where we have used a Clebsch-Gordan expansion for the product of two $D$-functions \cite{Zare} and the quantities in square brackets are Clebsch-Gordan coefficients. Interchanging the order of summation, we obtain
\begin{equation}\label{A4}
\hat{S}^{(1)}_{q_1} \hat{S}^{(1)}_{q_2} = \sum_k \cgc{1}{q_1}{1}{q_2}{k}{q_1+q_2} \sum_{p} \mathcal{D}^k_{p,q_1+q_2} (\hat{\Omega}) \sum_{p_1} \cgc{1}{p_1}{1}{p-p_1}{k}{p}  S_{p_1}^{(1)} S^{(1)}_{p-p_1},
\end{equation}
where $p=p_1+p_2$. The last sum  in this expression is, by definition, a spherical tensor product of $\hat{\bm{S}}$ with itself \cite{Zare}, so Eq. (\ref{A4}) takes the form
\begin{equation}\label{A5}
\hat{S}^{(1)}_{q_1} \hat{S}^{(1)}_{q_2} = \sum_k \cgc{1}{q_1}{1}{q_2}{k}{q_1+q_2} \sum_{p} \mathcal{D}^k_{p,q_1+q_2}(\hat{\Omega}) [\hat{\bm{S}} \otimes \hat{\bm{S}}]^{(k)}_p.
\end{equation}

Using the standard relations between the Cartesian and spherical tensor components of angular momentum operators \cite{Zare}, we get three important particular cases of Eq. (\ref{A5}) \cite{note2}:
\begin{align}\label{A6}\notag
\hat{S}_0^2 &= \hat{S}_z^2 = \sum_{k=0,2} \cgc{1}{0}{1}{0}{k}{0} \sum_p \mathcal{D}^k_{p,0}(\hat{\Omega}) [\hat{\bm{S}} \otimes \hat{\bm{S}}]^{(k)}_p, \\ \notag
\hat{S}_\pm^2 &= 2\hat{S}_{\pm 1}^2 = 2 \sum_{k=0,2} \cgc{1}{\pm1}{1}{\pm 1}{k}{\pm 2} \sum_p \mathcal{D}^k_{p,\pm 2}(\hat{\Omega}) [\hat{\bm{S}} \otimes \hat{\bm{S}}]^{(k)}_p, \\ 
\hat{S_\pm}\hat{S}_\mp &= -2\hat{S}^{(1)}_{\pm1} S^{(1)}_{\mp1} = - 2 \sum_{k=0,2} \cgc{1}{\pm1}{1}{\mp 1}{k}{0} \sum_p \mathcal{D}^k_{p0}(\hat{\Omega}) [\hat{\bm{S}} \otimes \hat{\bm{S}}]^{(k)}_p.
\end{align}
By combining these expressions with the definitions $\hat{S}_{x}= \frac{1}{2}( \hat{S}_+ + \hat{S}_-)$ and $\hat{S}_y = \frac{1}{2i}(\hat{S}_+ - \hat{S}_-)$ and neglecting the terms proportional to $\hat{S}^2$, we obtain Eq. (\ref{Hss_BF}) of the main text.

We note that the following SF tensor expression for the spin-rotation interaction (\ref{Hsr_BF})  
\begin{equation}
 H_\text{sr} = \bar{\gamma} \hat{\bm{N}} \cdot \hat{\bm{S}} + \sum_{p=-2}^2 \biggl{[}  {\textstyle \frac{1}{2}}  (\gamma_x-\gamma_y)  [\mathcal{D}^2_{p,2}(\hat{\Omega}) + \mathcal{D}^2_{p,-2}(\hat{\Omega})] +   {\textstyle \frac{1}{\sqrt{6}}}  (2\gamma_z-\gamma_x-\gamma_y) \mathcal{D}^2_{p,0}(\hat{\Omega}) \biggr{]} [\hat{\bm{N}} \otimes \hat{\bm{S}}]^{(2)}_p,
\end{equation}
where $\bar{\gamma}=\frac{1}{3}(\gamma_x+\gamma_y+\gamma_z)$, may be obtained by replacing the second spin operator in Eq. (\ref{A3}) with rotational angular momentum $\hat{\bm{N}}$ and retracing the steps outlined above. We note, however, that Eq. (\ref{Hsr_BF}) is simpler and hence more convenient for the practical purpose of evaluating the matrix elements in the fully uncoupled SF basis (\ref{expansion}).

\section*{Appendix B: Centrifugal distortion matrix elements}

Since the centrifugal distortion Hamiltonian (\ref{Hcd}) is spin-independent,  its matrix elements are diagonal in $S$ and $M_S$, so we omit these quantum numbers for the rest of this section. In addition, since the MF angular momentum operators do not couple states with different $N$ or $M_N$ (and independent of $M_N$), we can set $N=N'$ and $M_N=M_N'$ in all the expressions below.   The matrix elements of Eq. (\ref{Hcd}) in the spherical-top basis $|NM_NK_N\rangle$ can be evaluated by using the standard angular momentum algebra \cite{Zare} with the result for the first term
\begin{equation}\label{B1}
\langle NM_N K_N | -\Delta_N \hat{N}^4 | N M_N K_N' \rangle = -\Delta_N N^2(N+1)^2  \delta_{K_N K_N'},
\end{equation}
the second term
\begin{multline}\label{B2}
\langle NM_N K_N | -\Delta_{NK} \hat{N}^2 \hat{N}_x^2 | N M_N K_N' \rangle = -\frac{\Delta_{NK}}{4}N(N+1) \biggl{[} \lambda_\pm (N,K_N')\lambda_\pm(N,K_N'\pm1) \delta_{K_N,K_N'\pm 2} \\ +  2 [N(N+1) - K_N'^2] \delta_{K_N K_N'} \biggr{]},
\end{multline}
the third term
\begin{multline}\label{B3}
\langle NM_N K_N | -\Delta_{K} \hat{N}_x^4 | N M_N K_N' \rangle \\= -\frac{\Delta_{K}}{16} \biggl{[} \lambda_\pm (N,K_N')\lambda_\pm(N,K_N'\pm1) \lambda_\pm(N,K_N'\pm 2) \lambda_\pm(N,K_N'\pm 3) \delta_{K_N,K_N'\pm 4} \\ 
+ 2\lambda_\pm (N,K_N')\lambda_\pm(N,K_N'\pm1) [2N(N+1) - (K'\pm 2)^2 - K'^2] \delta_{K_N,K_N'\pm 2}  \\
+ \bigl{\{} 
 6[N(N+1)-K_N'^2]^2 - 4[N(N+1)-K_N'^2] + 6K_N'^2 \bigr{\}} \delta_{K_N,K_N'} \biggr{]}
\end{multline}
and, finally, the fourth term
\begin{multline}\label{B4}
\langle NM_N K_N | -2\delta_{N} \hat{N}^2 (\hat{N}_z^2 - N_y^2) | N M_N K_N' \rangle = -2 \delta_N N(N+1) \\ \times  \biggl{[} \frac{1}{4} \lambda_\pm(N,K_N') \lambda_\pm (N,K_N'\pm1) \delta_{K_N,K_N'\pm 2} 
 +   \bigl{\{}  K_N'^2 - \frac{1}{2} [N(N+1) - K_N'^2]   \bigl{\}} \delta_{K_N K_N'} \biggr{]}.
\end{multline}
The matrix element of the centrifugal distortion Hamiltonian can be obtained by summing Eqs. (\ref{B1}) through (\ref{B4}).

\section*{Appendix C: Angular expansion of the interaction potential}

Here, we derive the expressions for the radial expansion coefficients $V_{\mu\nu}(R)$ based on the {\it ab initio} PESs $V(R,\theta,\phi)$ calculated as described in Sec. II.
The expansion of the atom-molecule interaction PES in spherical harmonics (\ref{Vbf}) 
 \begin{equation}\label{C1}
V(R,\theta,\phi) = \sum_{\lambda=0}^{\lambda_\text{max}} \sum_{\mu \ge 0}^\lambda V_{\lambda\mu} (R) \frac{1}{1+\delta_{\mu 0}} \left(\frac{4\pi}{2\lambda+1}\right)^{1/2}  [Y_{\lambda\mu}(\theta,\phi) + (-1)^\mu Y_{\lambda,-\mu}(\theta,\phi)],
\end{equation}
may be rewritten in terms of real functions of $\theta$ and $\phi$ as
 \begin{equation}\label{C2}
V(R,\theta,\phi) = \sum_{\lambda=0}^{\lambda_\text{max}}  \sum_{\mu \ge 0}^\lambda V_{\lambda\mu} (R) \frac{2}{1+\delta_{\mu 0}} \left(\frac{2}{2\lambda+1}\right)^{1/2}  \Theta_{\lambda\mu}(\theta) \cos (\mu \phi),
\end{equation}
where $\Theta_{\lambda\mu}(\theta)$ is a normalized associated Legendre polynomial \cite{Zare}. A disadvantage of Eq. (\ref{C2}) is that the angular basis functions on the right-hand side are not orthonormal. It is more convenient to use an alternative expansion in terms of orthonormal angular basis functions 
\begin{equation}\label{C3}
V(R,\theta,\phi) = \sum_{\lambda=0}^{\lambda_\text{max}}  \sum_{\mu \ge 0}^\lambda \tilde{V}_{\lambda\mu}(R) \Theta_{\lambda\mu}(\theta) \left[ \frac{2}{(1+\delta_{\mu 0})\pi}\right]^{1/2}   \cos (\mu \phi).
\end{equation}
The expansion coefficients $ \tilde{V}_{\lambda\mu}(R) $ may be obtained by inverting this expression 
\begin{equation}\label{C4}
\tilde{V}_{\lambda\mu}(R)  =\left[ \frac{1}{2(1+\delta_{\mu 0})\pi}\right]^{1/2} \int_{0}^{2\pi} d\phi \int_0^{\pi} \sin\theta d\theta  V(R,\theta,\phi) \Theta_{\lambda\mu}(\theta)    \cos (\mu \phi).
\end{equation}
The relationship between the coefficients $\tilde{V}_{\lambda\mu}(R)$  and ${V}_{\lambda\mu}(R)$ can be obtained by comparing  Eqs. (\ref{C2}) and (\ref{C3}).  Using Eq. (\ref{C4}), we obtain the final  result
\begin{equation}\label{C5}
{V}_{\lambda\mu}(R)  = \frac{1}{2\pi} \left( \frac{2\lambda+1}{2}\right) ^{1/2} \int_{0}^{2\pi} d\phi \int_0^{\pi} \sin\theta d\theta  V(R,\theta,\phi) \Theta_{\lambda\mu}(\theta)    \cos (\mu \phi).
\end{equation}

\newpage

\begin{table}
\caption{Parameters of PES A}
\label{tab:pes}
\begin{center}
\begin{tabular}{lccccr}
\hline \hline
Pairs  & & & $A_{\alpha}$/eV  & $b_{\alpha}/a_0^{-1}$  & $C_{6,\alpha}/\text{eV}a_0^{6}$ \\ 
\hline
He-C         & & & 8.67689$\times 10^{-2} $ & 1.03161 & $-$1.30489$\times 10^{3}$ \\ 
He-H         & & & 24.90525                 & 1.66630 &    1.54137$\times 10^{2}$ \\ 
He-G$_{1,2}$ & & & 20.15632                 & 1.42113 &    3.93810$\times 10^{2}$ \\ 
He-G$_3$     & & & 1.07027$\times 10^{2} $  & 1.82350 &    1.99657$\times 10^{2}$ \\
\hline \hline 

\end{tabular}
\end{center}
\end{table}

\newpage

\begin{table}[t!]
\caption{Molecular constants of CH$_2(\tilde{X})$ and its isotopomers used to parametrize the asymptotic Hamiltonian (\ref{Hmol}). All the constants for CH$_2$ are taken from Ref. \cite{Bunker} and expressed in units of cm$^{-1}$. The constants for CHD and CD$_2$ are taken from Refs. \cite{CHD} and \cite{CD2}. The spin-rotation interaction has a negligible effect on spin relaxation cross sections for CHD and CD$_2$, thus the parameters  $\gamma_\alpha$ were set to zero.} 
\centering
\vspace{0.3cm}

\begin{tabular}{cccc}
\hline
\hline
Constant      & CH$_2$                 & CHD                  & CD$_2$                 \\
\hline
$A$           & $73.06$                & 55.5243              & 37.7869                \\
$B$           & $8.42$                 & 5.6784               & 4.2296                 \\
$C$           & $7.22$                 & 4.9735               & 3.6947                 \\
$D$           & $0.778$                & 0.7567               & 0.7765                 \\
$E$           & $0.0399$               & 0.0461               & 0.0406                 \\
$\gamma_{x}$  & $4.46\times 10^{-4}$   & ...                  & ...                    \\
$\gamma_{y}$  & $-4.106\times 10^{-3}$ & ...                  & ...                    \\
$\gamma_{z}$  & $-5.148\times 10^{-3}$ & ...                  & ...                    \\
$\Delta_N$    & $3.01\times 10^{-4}$   & $1.97\times 10^{-4}$ & $9.295\times 10^{-5}$  \\
$\Delta_{NK}$ & $-1.966\times 10^{-2}$ & $-7.8\times 10^{-3}$ & $-4.964\times 10^{-3}$ \\
$\Delta_K$    & $1.991049$             & 1.403                & 0.56022                \\
$\delta_N$    & $1.012\times 10^{-4}$  & $6.19\times 10^{-5}$ & $2.264\times 10^{-5}$  \\
\hline
\hline
\end{tabular}\begin{flushleft}
\end{flushleft}
\end{table}

\newpage

\begin{table}[t!]
\caption{Cross sections for spin relaxation calculated for different isotopomers of CH$_2$ vs the rotational energy splitting $\Delta E_{02}$ (in units of $\Delta E_{02}(\text{CH}_2$)). The cross sections are normalized to the He + CH$_2$ inelastic cross section ($\sigma_\text{inel}=1.51\times 10^{-4}$ \AA$^2$ for $E_C=0.1$ cm$^{-1}$ and $1.24\times 10^{-4}$ \AA$^2$ for $E_C=0.01$ cm$^{-1}$).  The values of collision energy (cm$^{-1}$) are given in parentheses.}
\centering
\vspace{0.3cm}

\begin{tabular}{cccc}
\hline
\hline
Molecule & $\Delta E_{02}$ & $\sigma_\text{inel}$(0.1) & $\sigma_\text{inel}$(0.01) \\
\hline
CH$_2$   & 1               & 1                         & 1                          \\
CHD      & 2.19            & 2.14                      & 2.04                       \\
CD$_2$   & 3.98            & 4.37                      & 3.87                       \\
\hline 
\hline
\end{tabular}\begin{flushleft}
\end{flushleft}
\end{table}

\newpage

%
%
%
%
%


\begin{thebibliography}{99}
\bibitem{note1}
The effective Hamiltonian (\ref{Hmol}) is appropriate for a rigid polyatomic molecule in a given vibronic state, and hence cannot be used to describe collision-induced transitions between different electronic or vibrational levels.
\bibitem{note2}
Note that the spin operators $\hat{S}_\pm=\hat{S}_x \pm i\hat{S}_y$  are defined in terms of Cartesian components of $\hat{\bm{S}}$, while the operators $\hat{S}^{(1)}_{\pm 1}$ are expressed in a spherical tensor basis \cite{Zare}: $\hat{S}^{(1)}_{\pm 1} = \mp (1/\sqrt{2}) \hat{S}_\pm$. 
\bibitem{cmshift}
Isotopic substitution causes a small shift in the origin of the MF coordinate system which should, in principle, be taken into account via a coordinate scaling transformation  \cite{ZH}. As we are mainly interested in the effect of rotational splitting on collision dynamics,  we did not apply the scaling transformation in this work.
\end{thebibliography}
\end{document}